\documentclass[12pt]{amsart}

\usepackage{array}
\usepackage{amsmath}
\usepackage{mathrsfs}
\usepackage{mathtools}
\usepackage{overpic}
\usepackage{graphicx}
\usepackage{float}
\usepackage[all]{xy}
\usepackage{pifont}
\usepackage{fancyhdr}
\usepackage{amsmath}
\usepackage{amscd}
\usepackage{amsfonts}
\usepackage{amssymb}
\usepackage{amsthm}
\usepackage{bm}
\usepackage{geometry}
\usepackage{setspace}
\usepackage{mathrsfs}
\usepackage{graphicx}

\usepackage{booktabs}
\usepackage[labelfont=bf]{caption}
\usepackage{caption}
\usepackage{enumitem}
\usepackage{subfigure}
\usepackage[colorlinks=false, hidelinks]{hyperref}
\usepackage{slashed}
\usepackage{tikz,tikz-cd}

\newtheorem{theorem}{Theorem}[section]

\newtheorem{corollary}[theorem]{Corollary}
\newtheorem{lemma}[theorem]{Lemma}

\newtheorem{proposition}[theorem]{Proposition}

\newtheorem{remark}[theorem]{Remark}
\theoremstyle{definition}
\newtheorem{definition}[theorem]{Definition}

\numberwithin{equation}{section}
\newcommand{\norm}[1]{\left\Vert#1\right\Vert}
\newcommand{\abs}[1]{\left\vert#1\right\vert}

\newcommand{\dt}{\mathrm{det}}
\newcommand{\R}{\mathbb R}
\newcommand{\Z}{\mathbb Z}

\newcommand{\C}{\mathbb C}
\newcommand{\CS}{\mathrm{CS}}

\newcommand{\tr}{\mathrm{tr}}
\newcommand{\Tr}{\mathrm{Tr}}

\newcommand\msv{\mathscr{V}}
\newcommand\mso{\mathscr{O}}

\newcommand\Hom{\mathrm{Hom}}

\newcommand\YM{\mathrm{YM}}

\newcommand\sse{\subseteq}

\newcommand\mbc{\mathbb{C}}

\newcommand\mca{\mathcal{A}}
\newcommand\mcl{\mathcal{L}}

\newcommand{\g}{\mathfrak{g}}

\newcommand{\ie}{\textit{i}.\textit{e}., }

\title{Quantization Commutes With Reduction of Chern-Simons Gauge Theory}
\author[G. Dai]{Geyang Dai}
\author[R. Liang]{Ruiming Liang}
\author[Y. Zhang]{Yang Zhang}

\address{Department of Mathematics, National University of Singapore} 
\address{Department of Mathematics, Peking University}
\address{Department of Mathematics, Rutgers University}
\email{e0983446@u.nus.edu}
\email{hiltonliang@stu.pku.edu.cn}
\email{yang.zhang18@rutgers.edu}
\date{}

\makeatletter
\renewcommand{\section}{\@startsection{section}{1}%
  \z@{.7\linespacing\@plus\linespacing}{.5\linespacing}%
  {\normalfont\large\bfseries}}
\renewcommand{\subsection}{\@startsection{subsection}{2}%
  \z@{.6\linespacing\@plus.7\linespacing}{.5\linespacing}%
  {\normalfont\normalsize\bfseries}}
\makeatother

\begin{document}

\begin{abstract}
We prove an infinite-dimensional version of ``quantization commutes with reduction" in the framework of geometric quantization of Chern-Simons gauge theory \cite{Axelrod1991GeometricQO}, focusing on the genus one case. The proof is complex-analytic and relies on the Atiyah-Bott stack and the Chern-Simons line bundle.
\end{abstract}

\maketitle

\tableofcontents

\section{Introduction}

``Quantization commutes with reduction" was first formulated by Guillemin-Sternberg in the setting of finite dimensional K\"ahler reduction \cite{Guillemin1982GeometricQA}. They considered a compact K\"ahler manifold $(M,\omega)$ with a Hamiltonian action of a compact Lie group $G$, together with a $G$-equivariant prequantum line bundle $\mathcal{L}$. 
Let $\mu$ be the moment map. If $G$ acts freely $\mu^{-1}(0)$, the K\"ahler quotient $M_G=M//G=\mu^{-1}(0)/G$ is a smooth K\"ahler manifold, and $\mathcal{L}$ descends to a prequantum line bundle $L_G\to M_G$.
They proved 
\begin{align}\label{Guillemin-Sterberg version}
H^0_G(M,\mathcal{L}^k)=H^0(M_G,L_G^k).
\end{align}
When $M_G$ is singular, this theorem was established in \cite{SjamarrSlice}.
A further advance was the formulation of the conjecture in terms of the index of $\mathrm{Spin}^c$ Dirac operators:
\begin{align*}
\mathrm{dim}Q(M,\mathcal{L})^G=\mathrm{dim}Q(M_G,L_G),
\end{align*}
where $Q(M,L)=\mathrm{Ker}(D_L)-\mathrm{Coker}(D_L)$ is a virtual $G$-representation.
Meinrenken \cite{Meinrenken1995SymplecticSA} proved this by symplectic cut. 
For singular reduction, see \cite{Meinrenken1997SINGULARRA}.
Tian–Zhang \cite{Tian1998AnAP} developed a direct analytic approach, which was later applied in non-compact settings (see, e.g., \cite{Ma2008GeometricQF}).

A natural direction is to extend the conjecture to infinite-dimensional geometric quantization.
Let $\Sigma$ be a Riemann surface and $G$ a compact, semi-simple, connected and simply connected Lie group.
Under this assumption, every principal $G$-bundle over $\Sigma$ is trivial as a smooth principal bundle.
Denote by $\mathcal{A}=\Omega^1(\Sigma,\g)$ the space of $G$-connections and by $\Sigma G$ the gauge group.
According to Atiyah-Bott \cite{Atiyah1971RiemannSA}, the gauge action makes $(\mathcal{A},\omega,\mu)$ a Hamiltonian $\Sigma G$-space where $\omega$ is a $\Sigma G$-invariant symplectic form and the moment map $\mu$ sends a connection to its curvature.
The prequantum line bundle can be described as the Chern-Simons line bundle $(\mathcal{L},h,\nabla)$ in \cite{Freed1995ClassicalCT}.

Via the infinite-dimensional GIT developed using the Yang-Mills flow, the K\"ahler reduction $M_G$, which is a possibly singular complex variety, parametrizes the S-equivalence classes of semi-stable $G_{\C}$-bundles over $\Sigma$.  $\mathcal{L}$ descends to a line bundle $\widetilde{\mathcal{L}}$ over $M_G$.
The infinite-dimensional analogue of \eqref{Guillemin-Sterberg version} then reads
\begin{align}\label{Infinite GIT}
H^0_{\Sigma G}(\mathcal{A},\mathcal{L}^k)=H^0(M_G,\widetilde{\mathcal{L}}^k).
\end{align}
The right-hand side is called the conformal block, 
whose dimension is famously given by the Verlinde formula (see e.g., \cite{Bismut1999SymplecticGA} in the context of geometric quantization).
The left-hand side, by contrast, remains much less understood because of its infinite-dimensional nature. 
Physically, it encodes the chiral Ward identities arising from the holomorphic factorization of the gauged Wess–Zumino–Witten model \cite{witten_verlinde_1993,Gawdzki1994SU2WT}.

In this paper, we aim to provide a rigorous analytic proof of \eqref{Infinite GIT}.
Substantial work has already been carried out in the algebraic–geometric framework.
By Teleman \cite{Teleman1998TheQC}, the Atiyah–Bott quotient stack $[\mathcal{A}/\Sigma G_{\C}]$ is the underlying complex-analytic stack of the moduli stack $\mathfrak{M}$ of $G_{\C}$-bundles over $\Sigma$. 
He also
 proved the quantization conjecture and the vanishing of higher cohomology groups of the line bundle $\mathcal{L}$ on the algebraic stack $\mathfrak{M}$. The case of global sections $H^0$ goes back to \cite{Beauville1993ConformalBA} for $G=SL_n$, and to \cite{Kumar1994InfiniteGA,Laszlo1997TheLB} in general $G$. 

We focus mainly on the genus one case for several reasons. In higher genus, the quantization conjecture has been proved (\cite{Kumar1994InfiniteGA,Laszlo1997TheLB,Beauville1993ConformalBA}), essentially because the complement of stable points has codimension at least $2$ \cite{Atiyah1983TheYE}.
This also leads to a parallel complex–analytic proof of \eqref{Infinite GIT}, although care is needed in describing the local model of the singularity via the Kuranishi map (see, e.g., \cite{Huebschmann1995TheSO}).
 
 By contrast, in the genus one case, the stable locus is empty, so a suitable replacement is required. Moreover, the genus one coarse moduli admits a global quotient model $M_G=(E\otimes \Lambda)/W$, where $E$ is the elliptic curve, $\Lambda$ is the co-root lattice, and $W$ is the Weyl group.
The genus one conformal block  can then be realized as Weyl group invariant combinations of theta functions, as in \cite{Looijenga1976RootSA}.

Our main result is the following:

\begin{theorem}\label{QR=0}

Let $E$ be an elliptic curve and $G$ be a compact, semisimple, connected, and simply connected Lie group.
Let $\mathcal{L}\to \mca$ be the Chern-Simons line bundle.
Set $\mathcal{H}_k(G)= H^0_{\Sigma G}(\mathcal{A}, \mathcal{L}^k)$ and $V_k(G)= H^0(M_G, \widetilde{\mathcal{L}}^k)$ as defined in Definition \ref{Conformal Block}.
Then
\begin{align}
\mathcal{H}_k(G) = V_k(G).
\end{align}
 \end{theorem}
 
To prove this, we adapt the observation of \cite{SjamarrSlice} to the infinite-dimensional setting and show that the norm of any equivariant holomorphic section is non-decreasing along the Yang–Mills flow (see Lemma \ref{IncreasingNorm}).
Since the stable locus is empty, we use the locus of poly-stable and regular points as a replacement and prove that it contains a Zariski open subset. Hence, it captures the pushing-down geometry on the coarse moduli without loss of information.

During explorations,
we also establish an isometry between the determinant (resp.\ Pfaffian) line bundle and the Chern–Simons line bundle, viewed as $\Sigma G$–equivariant prequantum line bundles over the space of connections $\mca$ (see Theorem \ref{detpf}). 
This extends \cite{Ramadas1989SomeCO}, which proved the isometry only after reduction.
In the genus one case, the index of the family of Dirac operators is always zero.
Consequently, the canonical section of the determinant (resp.\ Pfaffian) line bundle is well defined. 
 As an application, we show that the $\zeta$–regularized determinant of the Laplacian increases along the Yang–Mills flow (see Proposition \ref{regularized det increases}).

\subsubsection*{\textbf{Organization of the paper}}
\begin{itemize}

\item In Section \ref{GQ of CS}, we review the geometric setup of
quantization of Chern-Simons gauge theory. We also give an isometry between the determinant (resp.\ Pfaffian) line bundle and the Chern-Simons line bundle as $\Sigma G$-equivariant prequantum line bundles over $\mathcal{A}$.
\item In Section \ref{GIT of Yangmills flow},
we review some results about the Yang-Mills flow on Riemann surfaces and the generalized Narasimhan-Seshadri-Ramanathan Theorem.
\item In Section \ref{Sec:Proof}, we prove the main theorem using the Yang-Mills flow and the extension method.
\item In Section \ref{Sec:Further Discussions}, we give some applications and discussions with related work from representation theory.

\end{itemize}

\section{Geometric Quantization of Chern-Simons Gauge theory}\label{GQ of CS}

This section reviews the geometric setup for geometric quantization of Chern-Simons gauge theory \cite{Axelrod1991GeometricQO}.
Throughout this paper, $G$ is a compact, semi-simple, connected and simply connected Lie group, $\mathcal{A}=\Omega^1(\Sigma,\mathfrak{g})$ denotes the space of $\mathfrak{g}$-value connections, $\Sigma G$ denotes the gauge group and we always assume the connections are of Sobolev class $W^{1,2}$ and the gauge groups are of Sobolev class $W^{2,2}$.

\subsection{Hamiltonian Geometry of the Space of Connections}

Let $\mathrm{ad}\g$ be the adjoint representation of $\g$.
The tangent space of $\mathcal{A}$ at $A$ is $T_A(\mathcal{A})=\Omega^1(\Sigma,\mathrm{ad}\g)$.
We denote by
\begin{align*}
d_A=d+[A,\cdot]:\Omega^i(\Sigma,\mathrm{ad}\g)\to \Omega^{i+1}(\Sigma,\mathrm{ad}\g) 
\end{align*}
the associated covariant derivative, and by
\begin{align*}
   F_A=dA+\frac{1}{2}[A,A]\in \Omega^2(\Sigma,\g) 
\end{align*}
its curvature.
The gauge group $\Sigma G$ acts on $\mathcal{A}=\Omega^1(\Sigma,\mathfrak{g})$ by the gauge transformations
\begin{align*}
gA=gAg^{-1}-dgg^{-1}.   
\end{align*}
Here, $\Sigma \g$ denotes the Lie algebra of the gauge group $\Sigma G$. Given
$\xi\in \Sigma \g$, 
the fundamental vector field  $\xi^{\#}$ on $\mathcal{A}$ is given by
\begin{align}\label{fundamental vector field}
 \xi^{\#}_A=-d_A\xi.   
\end{align}  
Let $\langle\cdot,\cdot \rangle$ be the normalized invariant bilinear form on $\g$ such that the longest roots have square length $2$.
The minimal symplectic form
$\omega$ is given by
\begin{align*}
\omega_A(a_1,a_2)=\int_{\Sigma}\langle a_1\wedge a_2\rangle.
\end{align*}
It is straightforward to check that it is closed and $\Sigma G$-invariant.

\begin{proposition}[{\cite{Atiyah1983TheYE}}]\label{Atiyah-Bott reduction}
The gauge group $\Sigma G$-action is Hamiltonian, and the corresponding moment map 
$
\mu:\Omega^1(\Sigma,\g)\to \Omega^0(\Sigma,\g)^*
$
is given by
\begin{align*}
\langle \mu,\xi\rangle_A=\int_{\Sigma}\langle F_A\wedge \xi\rangle.
\end{align*}
We call 
$(\mathcal{A},\omega,\mu)$ a Hamiltonian $\Sigma G$-space.
\end{proposition}

Given a complex structure on $\Sigma$, the Hodge star operator $*$ on $\Omega^1(\Sigma)$ satisfies $*^2=-\mathrm{Id}$ and extends to $\mathcal{A}=\Omega^1(\Sigma,\mathfrak{g})$, endowing $\mathcal{A}$ with a complex structure. With this structure, the real vector space $\mathcal{A}$ together with $*$, is identified with $\Omega^{0,1}(\Sigma,\mathfrak{g}_{\mathbb{C}})$. 
A functional $f:\mathcal{A}\to\mathbb{C}$ is holomorphic if, for every $A\in\mathcal{A}$ and $v\in T_A(\mathcal{A})$, one has $df_A(*v)=i\, d f_A(v)$.

Now 
$(\mathcal{A},\omega)$ has the K\"{a}hler structure with the compatible metric 
\begin{align*}
  \langle a,b\rangle_A:=\omega_A(a,*b)=\int_{\Sigma}\langle a\wedge*b\rangle.
\end{align*}
On the Lie-algebra level,
$\Sigma \g_{\C}$-action on $\mathcal{A}$ is given by
\begin{align*}
d_A(\xi+i\eta)=d_A\xi+*d_A\eta.
\end{align*}
Given $\xi\in \Sigma \g$, 
the fundamental vector field of $i\xi$ on $\mathcal{A}$ is
\begin{align*}
(i\xi)^{\#}=*\xi^{\#}.
\end{align*}
Using the identification $(\mathcal{A}=\Omega^1(\Sigma,\g),*) \cong (\Omega^{0,1}(\Sigma,\g_{\C}),i)$, the $\Sigma G_{\C}$-action on $\mathcal{A}$ is the complexification of the gauge action. For complex-analytic purposes, we henceforth regard $\mathcal{A}$ as $\Omega^{0,1}(\Sigma,\g_{\C})$. Under this identification, the $\Sigma G_{\C}$-action on $\Omega^{0,1}(\Sigma,\g_{\C})$ is given by
 \begin{align*}
    gA=gAg^{-1}-\bar{\partial}gg^{-1}.
 \end{align*}
The Cauchy-Riemann operator coupled with connection $A$ is 
\begin{align*}
\bar{\partial}_A=\bar{\partial}+\mathrm{ad}_A:\Omega^0(\Sigma,\mathrm{ad}\g_{\C})\to \Omega^{0,1}(\Sigma,\mathrm{ad}\g_{\C}).
\end{align*}
The operator
 $\bar{\partial}_A$ defines a holomorphic structure on the adjoint bundle $\mathrm{ad}\g_{\C}$.
Two connections $A$ and $A'$ define equivalent holomorphic structures if and only if $A\sim A'$ under the $\Sigma G_{\C}$-action.
Denote by $V_A$ the corresponding holomorphic adjoint bundle determined by $\bar{\partial}_A$.
Likewise, the connection $A$ determines a holomorphic principal $G_{\C}$-bundle $P_A\to \Sigma$, whose adjoint bundle is $V_A=\mathrm{ad}(P_A)$.
Let $H^i(\Sigma,V_A)$ be the sheaf cohomology of $V_A$, we have the following
\begin{equation*}
H^0(\Sigma,V_A)=\mathrm{Ker}(\bar{\partial}_A),\ H^1(\Sigma,V_A)=\mathrm{Ker}(\bar{\partial}_A^*).
\end{equation*}

\subsection{Chern-Simons Line Bundle and Quantization before Reduction}

\subsubsection{Chern-Simons line bundle}

We review the construction of Chern–Simons line bundle \cite{witten_verlinde_1993,Freed1995ClassicalCT}.

Chern–Simons line bundle $\mathcal{L}=\mathcal{A}\times\C \to \mathcal{A}$ carries a nontrivial $\Sigma G$-action defined by the gauged Wess–Zumino–Witten cocycle, which measures the change of the Chern–Simons functional under gauge transformations on a threefold with boundary $\Sigma$.

Let $B$ be a threefold with boundary $\partial B=\Sigma$. A pair $(\widetilde{A},\widetilde{g}) \in \Omega^1(B,\g)\times W^{2,2}(B,G)$ is called an extension of $(A,g)\in \mathcal{A}\times \Sigma G$ if
$
(\widetilde{A},\widetilde{g})|_{\partial B}=(A,g).
$
The gauged Wess–Zumino–Witten cocycle $W:\Sigma G\times \mathcal{A}\to \R/2\pi \Z$ is defined by:
\begin{align*}
W(g,A)=\int_B \big(\CS(\widetilde{A})-\CS(\widetilde{g}\,\widetilde{A})\big),
\end{align*}
where the Chern–Simons $3$–form is
\begin{align*}
    \CS(A)= \frac{1}{4\pi}\Big\langle A\wedge dA+\frac{2}{3}\,A\wedge A\wedge A\Big\rangle.   
\end{align*}
It is independent of the choice of extension $(B,\widetilde{A},\widetilde{g})$, and one checks that
\begin{align*}
 e^{iW(gh,A)}=e^{iW(g,hA)}\cdot e^{iW(h,A)}.       
\end{align*}
Consequently, it defines a $\Sigma G$–equivariant line bundle as follows.

\begin{definition}
Chern–Simons line bundle is a $\Sigma G$–equivariant line bundle $\mathcal{L}=\mathcal{A}\times \C \to \mathcal{A}$, where the $\Sigma G$–action is given, for $(A,z)\in \mathcal{A}\times \C$ and $g\in \Sigma G$, by
\begin{align*}
g(A,z)=(gA,e^{iW(g,A)}z).    
\end{align*}
Let $h$ be the trivial metric.
The connection $\nabla=d-\sqrt{-1}\theta$,
where $\theta\in \Omega^1(\mathcal{A},\R) $ is defined as, for
$A\in \mathcal{A}$ and $a\in T_A\mathcal{A}$,
\begin{align*}
\theta_A(a)=\frac{1}{2}\int_{\Sigma}\langle A\wedge a\rangle. 
\end{align*}
\end{definition}

\begin{proposition}[{\cite[Proposition 3.17]{Freed1995ClassicalCT}}]\label{UCS is equivariant prequantumline}

$(\mathcal{L},h,\nabla)$ is a $\Sigma G$-equivariant prequantum line bundle over the Hamiltonian $\Sigma G$-space $(\mathcal{A},\omega,\mu)$.

\end{proposition}

\subsubsection{Quantization before Reduction}

From physical perspective, 
the space of quantization before reduction $\mathcal{H}_k(G)$ consists of solutions to the chiral Ward identities arising in the holomorphic factorization of the gauged Wess-Zumino-Witten model.
Its dimension can be evaluated formally via path integrals, yielding the Verlinde formula before reduction.

When the complex structure of $\Sigma$ is given, 
the $(1,0)$-part of the connection $\nabla$ gives $\mathcal{L}$ a holomorphic structure. 
 The $\bar{\partial}$-operator 
$\bar{\partial}_{\mathcal{L}}$
is defined by $\bar{\partial}-\sqrt{-1}\theta^{1,0}$,
where the connection form $\theta^{1,0}$ is given by,
for 
$a\in (T_A\mathcal{A})_{\C}$,
\begin{align*}
\theta^{1,0}_A(a)=\frac{1}{2}\int_{\Sigma}\langle A\wedge a^{1,0}\rangle. 
\end{align*}
 Given $\chi:\mathcal{A}\to \C$, $\chi$ is a holomorphic section of $\mathcal{L}$ if
\begin{align*}
\bar{\partial}_{\mathcal{L}}\chi=0 .
\end{align*}
Equivalently,
$\chi$ can be written as
$
\chi(A)=\exp{(\frac{\sqrt{-1}}{2}K(A))}\widehat{\chi}(A) 
$,
where $\widehat{\chi}$ is a holomorphic functional of $\mathcal{A}$ and $K$ is the K\"ahler potential of $(\mca,\omega)$.

\begin{definition}
The space of
\emph{quantization before reduction} is the $\C$-vector space
\begin{align*}
    \mathcal{H}_k(G)=H^0_{\Sigma G}(\mathcal{A},\mathcal{L}^k).
\end{align*}
\end{definition}

There is a natural extension of $\Sigma G_{\C}$-action on $\mathcal{L}$.
On the Lie algebra level, the $\Sigma \g_{\C}$-action on sections can be described as the same as the finite dimensional case \cite[§5]{Guillemin1982GeometricQA}.
Given $\xi\in \Sigma \g$ and $\xi^{\#}$
the corresponding fundamental vector field \eqref{fundamental vector field},
 the section $s:\mathcal{A}\to \C$, by the definition of the moment map,
\begin{align}{\label{momemt map rel}}
\xi \cdot s=\nabla_{\xi^{\#}}s-2\pi\sqrt{-1}\langle \mu,\xi \rangle s.
\end{align}
When $s$ is a holomorphic section, 
\begin{align*}
(i\xi)\cdot s=\sqrt{-1}\xi\cdot s.
\end{align*}
Since $(i\xi)^{\#}=*\xi^{\#}$,
$\nabla_{(i\xi)^{\#}}s=\nabla_{*\xi^{\#}}s$.
$\xi^{\#}-\sqrt{-1}*\xi^{\#}$ is anti-holomorphic and $s$ is holomorphic, so we have 
\begin{align*}
\nabla_{\xi^{\#}-\sqrt{-1}(i\xi)^{\#}}s=\nabla_{\xi^{\#}-\sqrt{-1}*\xi^{\#}}s=0.
\end{align*}
By \eqref{momemt map rel},
\begin{align}\label{ImageAction}
(i\xi)\cdot s=-\nabla_{(i\xi)^{\#}}s+2\pi\langle \mu,\xi\rangle s.
\end{align}
For general section $s'=fs$, the action is given by
\begin{align*}
    (i\xi)\cdot (fs)=(*\xi^{\#}) f+\sqrt{-1}f(\xi\cdot s).
\end{align*}
Automatically, we have
\begin{align*}
    \mathcal{H}_k(G)=H^0_{\Sigma G}(\mathcal{A},\mathcal{L}^k)
=H^0_{\Sigma G_{\C}}(\mathcal{A},\mathcal{L}^k).
\end{align*}

Next,
one of the aims is to show that $\mathcal{H}_k(G)$ is at least nonempty for some $k\in \mathbb{Z}$. Even in the finite-dimensional setting,
this fact is nontrivial (see \cite[Theorem 5.6]{Guillemin1982GeometricQA}).

\subsection{Determinant Line bundle and Chern-Simons Line Bundle}

We apply the gauge-equivariant family index theorem to identify the determinant line bundle with an appropriate power of the Chern–Simons line bundle. 
In the genus one case, the canonical section defined via the $\zeta$-regularized determinant of the Laplacian yields a holomorphic $\Sigma G$-equivariant section.

The construction of the determinant line bundle, together with the Quillen metric and the Bismut–Freed connection, is well established (see \cite{Bismut1986TheAO,Bismut2}). The original construction of the determinant line bundle for Cauchy–Riemann operators, equivalently for the space of connections in our setting, goes back to \cite{Quillen1985DeterminantsOC,Atiyah1984DiracOC}. The theory of Pfaffian line bundles in the presence of real structures and their relation to physical anomalies are discussed in \cite{freed1987determinant,Freed1986DeterminantsTA}.
 
We first recall the level of a representation.
For finite-dimensional representation $\rho:\g \to \mathfrak{gl}(V)$.
The Dynkin index $d_V$ is the scalar defined by
\begin{align*}
  \tr(\rho(x)\rho(y))=d_V\langle x,y\rangle. 
\end{align*}
In particular, the adjoint representation has level $d_{\mathrm{ad}}=2h^{\vee}$, where $h^{\vee}$ is the dual Coxeter number of $\g$.

The rest of this section is to prove the following theorem,
which can be viewed as the gauge–equivariant analog of the determinant line bundle construction arising from family index theory over the space of connections.


It is a classical result \cite{Atiyah1971RiemannSA} that the square root of the canonical line bundle $\sqrt{K}$ defines a spin structure on the Riemann surface $\Sigma$.

\begin{theorem}{\label{detpf}}

Fix a spin structure $\sqrt{K}$ and a K\"ahler metric on $\Sigma$. For any finite-dimensional unitary representation $V$ of $G$, 
one can construct the determinant line bundle
$\mathrm{DET}(V)$ equipped with the Quillen metric $g^{\mathrm{DET}(V)}$ and the Bismut-Freed connection $\nabla^{\mathrm{DET}(V)}$ (for explicit constructions, see \ref{definition of determinant line bundle}), which has the following properties:
\begin{itemize}
    \item It is a $\Sigma G$-equivariant  prequantum line bundle of level $d_V$ over the 
    Hamiltonian $\Sigma G$-space $(\mathcal{A},\omega,\mu)$. The holomorphic structure is given by the $(1,0)$-part of the Bismut-Freed connection.
    \item It is isomorphic to the Chern-Simons line bundle $(\mathcal{L},h,\nabla)^{d_V}$.
    \item In the genus one case, the index of the family of Dirac operators is always zero. We can define the canonical section $\det_V\in H^0_{\Sigma G}(\mathcal{A},\mathrm{DET}(V))$, which has the property 
    \begin{align*}
    \norm{\mathrm{det}_V(A)}^2_{g^{\mathrm{DET}(V)}}=\mathrm{det}_{\zeta}(\slashed\partial_A^{V,*}\slashed\partial^V_A),
    \end{align*}
    where $\slashed\partial_A^V$ is the Dirac operator twisted by vector potentials $A$ on $V$ and $\slashed\partial_A^{V,*}$ is the adjoint.
    $\dt_{\zeta}$ is called the $\zeta$-function regularized determinant of the Laplacian.
\end{itemize}

\end{theorem}
In particular, when $V$ is a real representation, the determinant line bundle admits a canonical square root, known as the Pfaffian line bundle, which satisfies properties analogous to those above and has level $d_V/2$.
 
\subsubsection{Determinant line bundle as $\Sigma G$-equivariant prequantum line bundle}\label{definition of determinant line bundle}

In order to define the determinant line bundle over $\mathcal{A}$, we fix the K\"{a}hler metric of $\Sigma$, the spin structure $\sqrt{K}$, and the unitary representation  $V$ of $G$. With this geometric information, $\mathcal{A}$ parametrizes the family of Dirac operators coupled with varying vector potentials. Following \cite{freed1987determinant}, we recall the construction of the determinant line bundle equipped with the Quillen metric and the Bismut–Freed connection $(\mathrm{DET}(V),g^{\mathrm{DET}(V)},\nabla^{\mathrm{DET}(V)})$, and prove that it is $\Sigma G$-equivariant over $\mathcal{A}$. In the genus one case, we provide the construction of the canonical section and prove that it is holomorphic and $\Sigma G$-equivariant.
  
Given a finite-dimensional unitary representation $ V\in \text{Rep}(G)$, there is an associated vector bundle of the unitary representation. Since it is trivial as a smooth principal bundle, we still denote it by $V$. There is the splitting of the bundle connection 
$
\nabla_A=\nabla_A^{1,0}+\nabla_A^{0,1},
$
where $\nabla_A^{0,1}$ serves as the $\bar{\partial}$-operator defining the holomorphic structure. The corresponding holomorphic vector bundle 
is denoted by $V_A$. The Dirac operator coupled with the vector potential is
\begin{align*}
 \slashed {\partial}_A^V:\Omega^0(\Sigma,V\otimes \sqrt{K})\to \Omega^{0,1}(\Sigma,V\otimes \sqrt{K}).
\end{align*}
and the adjoint,
\begin{align*}
 \slashed {\partial}_A^{V,*}:\Omega^{0,1}(\Sigma,V\otimes\sqrt{K}) \to \Omega^{0}(\Sigma,V\otimes \sqrt{K}).    
\end{align*}
The kernel and the co-kernel of the Dirac operator are given by the sheaf cohomology groups
\begin{align*}
\mathrm{Ker}(\slashed {\partial}_A^V)=H^0(\Sigma,V_A\otimes \sqrt{K}) \quad
\mathrm{Ker}(\slashed {\partial}_A^{V,*})=H^1(\Sigma,V_A\otimes \sqrt{K}).
\end{align*}
The index is given by
$\mathrm{Ind}(\slashed {\partial}_A^V)=\dim H^0(\Sigma,V_A\otimes \sqrt{K})-\dim H^1(\Sigma,V_A\otimes \sqrt{K})$. Since $V_A$ is topologically trivial, the Riemann--Roch theorem implies that
\begin{align*}
 \mathrm{Ind}(\slashed {\partial}_A^V)=\chi(\Sigma,V_A\otimes \sqrt{K} ) =\mathrm{deg}(V)+r(V)(1-g(\Sigma)).
\end{align*}
Here, $\mathrm{deg}(V)$ is the degree of the vector bundle $V$, which is zero here since $V$ is smoothly trivial, $r(V)$ is the dimension of the representation $V$ and $g(\Sigma)$ is the genus of $\Sigma$. In particular, in the genus one case, the index is always zero. Spinor bundles are defined as $\mathcal{H}_{+}=\Omega^0(\Sigma,V\otimes \sqrt{K}),\mathcal{H}_{-}=\Omega^{0,1}(\Sigma,V\otimes \sqrt{K})$ which are equipped with metrics induced by the Hermitian structure of $V$. For simplicity, we write
$D_A=\slashed{\partial}_A^{V}$ and $(D_A)^*=\slashed{\partial}_A^{V,*}$. Let $\Delta_A^V=D_A^*D_A$. 
Given  $ g \in \Sigma G $ , its action on the bundle via the unitary representation yields a  $ U(V) $ -valued gauge transformation. Since this action is unitary, we have  $ g^* = g^{-1} $ .

The following operators transform as
\begin{align*}
D_{gA}=gD_Ag^{-1}, \quad D_{gA}^*=gD_A^*g^{-1},\quad \Delta_{gA}^V=g\Delta_A^Vg^{-1}.
\end{align*}
The spectrum of the Laplacian $\Delta_A^V$ is gauge-invariant.

(1) The construction of $\mathrm{DET}(V)$ and the $\Sigma G$-equivariance.
    
We define the determinant line bundle using the spectrum cover $U^{(a)}=\{A\in \mathcal{A}|a\notin \mathrm{spec}(\Delta_A^V) \}$ and let $\mathcal{H}_{+}^{(a)}$ be 
the sum of eigen-spaces for eigen-values less than $a$ and $\mathcal{H}_{-}^{(a)}$ be the analogous sum for the Laplacian $D_AD_A^*$.  The determinant line bundle locally is
$\mathcal{L}^{(a)} \to U^{(a)}$ by $\mathcal{L}^{(a)}=(\det\mathcal{H}_{+}^{(a)})^*\otimes (\dt \mathcal{H}_{-}^{(a)})$.

Since the gauge transformation $\Sigma G$ preserves the spectrum of the Laplacian, the $\Sigma G$-action can be restricted to $U^{(a)}$. Moreover, for $g\in \Sigma G$ and $\psi \in \mathcal{H}_{\pm}^{(a)}|_A$, we have $g\psi\in \mathcal{H}_{\pm}^{(a)}|_{gA}$. This induces a $\Sigma G$-action on $\mathcal{L}^{(a)}\to U^{(a)}$. For $a<b$, let $\mathcal{H}_{\pm}^{(a,b)}$ be be the span of eigenspaces with eigenvalues in $(a,b)$. This forms a smooth vector bundle over $U^{(a)}\cap U^{(b)}$. $D^{(a,b)}:\mathcal{H}_+^{(a,b)}\to \mathcal{H}_-^{(a,b)}$ fiberwise is an isomorphism, and thus defines a nowhere-vanishing section $\dt D^{(a,b)}$ of $\mathcal{L}^{(a,b)}$. This yields a canonical isomorphism
\begin{align*}
 \mathcal{L}^{(b)} \to \mathcal{L}^{(a)}\otimes \mathcal{L}^{(a,b)},\quad
 s\to s\otimes \dt D^{(a,b)}.
\end{align*}
Since $D^{(a,b)}$ is $\Sigma G$-equivariant, this isomorphism is also $\Sigma G$-equivariant. Finally, $\mathrm{DET}(V)\to \mathcal{A}$ is defined by patching $\mathcal{L}^{(a)}$ using the $\Sigma G$ equivariant isomorphism above, thus is $\Sigma G$-equivariant.

(2) In the genus one case, the canonical section $\dt_V$ is holomorphic and $\Sigma G$-equivariant.

The index of $D$ is always zero, so we have $\dim H_{+}^{(a)}=\dim H_{-}^{(a)}$. Hence, the operator $D^{(a)}:H_{+}^{(a)}\to H_{-}^{(a)}$ defines a canonical non-vanishing section $\dt D^{(a)}:\dt H_{+}^{(a)}\to \dt H_{-}^{(a)}$. These sections satisfy the compatibility condition $\dt D^{(a)}\otimes \dt D^{(a,b)}=\dt D^{(b)}$ and therefore glue to a global section $\dt_V$.
From the above, the canonical section $\dt_V$ is $\Sigma G$-equivariant. It is holomorphic since the $\bar{\partial}$ complex $D:H^+\to H^-$ varies holomorphically over $\mca$.

(3) Quillen metric $g^{\mathrm{DET}(V)}$ is $\Sigma G$-invariant.

The bundles $\mathcal{H}_{\pm}^{(a)}$ inherit metrics from $\mathcal{H}_{\pm}$, denoted $g^{(a)}_{\pm}$. Since the gauge transformation is unitary, $g_{\pm}^{(a)}$ is $\Sigma G$-invariant, and so is the induced metric $g^{(a)}$ on $\mathcal{L}^{(a)}$. However, the collection $\{g^{(a)}\}$ does not glue to a global metric on $\mathrm{DET}(V)$. To correct this, set 
\begin{align*}
\overline{g}^{(a)}=g^{(a)}\cdot \dt_{\zeta}(\Delta^V)_{>a},
\end{align*}
where  $\dt_{\zeta}(\Delta^V)_{>a}$ is the regularized product of eigenvalues of the Laplacian $\Delta^V$ greater than $a$.

Here $\zeta^{(a)}(s)=\tr((\Delta^V|_{\lambda>a})^{-s})$ is holomorphic if the real part $\Re(s)$ is sufficiently large, and has a meromorphic continuation to $\C$ which is holomorphic at $s=0$. The regularized determinant is defined as
\begin{align*}
\dt_{\zeta}(\Delta^V)_{>a}:=\mathrm{exp}(-\zeta^{(a)'}(0)).  
\end{align*}
It satisfies $\dt_{\zeta}(\Delta^V)_{>a}=\dt_{\zeta}(\Delta^V)_{>b}(\prod_{a<\lambda<b}\lambda)$, which ensures that $\overline{g}^{(a)}$ and $\overline{g}^{(b)}$ agree on the overlap. Thus, they patch to a global metric $g^{\mathrm{DET}(V)}$, known as the Quillen metric.

Furthermore, since the spectrum of the Laplacian is $\Sigma G$-invariant, the regularized determinant is $\Sigma G$-invariant. Combined with the $\Sigma G$-invariance of $g^{(a)}$, it follows that $\overline{g}^{(a)}$ is $\Sigma G$-invariant, and hence the Quillen metric is $\Sigma G$-invariant.

(4) Bismut-Freed connection $\nabla^{\mathrm{DET}(V)}$ is $\Sigma G$-invariant.

Define the connection $1$-form $\theta^{(a)}$ of $\mathcal{L}^{(a)}$ as
$
\theta^{(a)}=\tr(dDD^{-1})_{>a}
$.
Again we use $\zeta$-function to define the trace above 
$$
w^{(a)}(s)=\tr((DD^*)^{-s}dDD^{-1})_{>a}.
$$
The $1$-form $w^{(a)}(s)$ is holomorphic for $\Re(s)$ sufficiently large and extends meromorphically to $\C$ with a simple pole at $s=0$. Furthermore, the real part of the residue at $s=0$ is $-\frac{1}{2}d\zeta^{(a)}(0)$. 
Define 
$$\theta^{(a)}=\tr(dDD^{-1})_{>a}:=(sw^{(a)}(s))'(0),$$
which satisfies
$
\Re(\theta^{(a)})=-\frac{1}{2}d(\zeta^{(a)})'(0).
$
On overlaps $U^{(a)}\cap U^{(b)}$ with $a<b$, we have
$$
\theta^{(a)}=\theta^{(b)}+\tr(dDD^{-1})_{(a<\lambda<b)},
$$
 where $\tr(dDD^{-1})_{(a<\lambda<b)}=d(\dt D^{(a,b)})(\dt D^{(a,b)})^{-1}$. Thus the connection form can be patched together to the Bismut-Freed connection $\nabla^{\mathrm{DET}(V)}$. It is unitary for the Quillen metric. Finally, the $\Sigma G$-invariance of the Bismut-Freed connection follows from $D\to gDg^{-1}$. Hence $\tr(dDD^{-1})$ is invariant under gauge transformations.

\subsubsection{Universal bundles and equivariant curvature}\label{universal bundles}

We define the $\Sigma G$-equivariant version of universal bundle and universal connection, and use it to compute the equivariant curvature of the Bismut-Freed connection. 

Following \cite{Atiyah1984DiracOC}, but working directly in the $\Sigma G$-equivariant setting rather than passing to the quotient, we introduce the universal bundle $(\mathcal{E},u) \to \Sigma \times \mathcal{A}$, where $\mathcal{E}$ is a $\Sigma G$-equivariant principal $G$-bundle and $u$ is a $\Sigma G$-invariant $G$-connection, referred to as the universal connection.

The universal bundle is defined as $\mathcal{E}:=Q\times \mathcal{A}\to \Sigma \times \mathcal{A}$, where $Q=\Sigma \times G$ is the trivial $G$-bundle over $\Sigma$. 
The gauge group $\Sigma G$ acts on $Q\times \mathcal{A}$ by $(q,A)\mapsto (g(q),gA)$. This $\Sigma G$-action is free and commutes with the right $G$-action, so $\mathcal{E}$ is a $\Sigma G$-equivariant $G$-bundle over $\Sigma \times \mathcal{A}$.

The next step is to define a metric $g_{\mathcal{E}}$ on $\mathcal{E}$. 
To construct this metric, we use the bi-invariant metric on $G$ induced by the minimal form $\langle\cdot,\cdot \rangle$ on $\g$, together with the K\"{a}hler metric on $\Sigma$.
The connection $A\in \mathcal{A}$ induces a splitting of the tangent bundle $TQ$ into $VQ\oplus HQ$, where $HQ$ is the horizontal component. The metric $g(A)$ on $Q$ is defined using the metric on $\Sigma$ and the metric on $G$ given by the minimal form. From the construction, $g(A)$ is $G$-invariant. The metric $g_{\mathcal{E}}$ on $\mathcal{E}=Q\times \mathcal{A}$ combines the bi-invariant metric on $G$ with the K\"ahler metric on $\mathcal{A}$.
This metric is invariant under both the $G$-action and the $\Sigma G$-action, which allows us to define the orthogonal complement to the $G$-orbits and therefore obtain a $\Sigma G$-invariant $G$-connection $u$ on the universal $\Sigma G$-equivariant $G$-bundle $\mathcal{E}\to \Sigma \times \mathcal{A}$.

Let $\mathcal{E}_{\C}=\Sigma\times \mathcal{A}\times G_{\C}$ be the trivial principal $G_{\C}$-bundle over $\Sigma\times \mathcal{A}$. With respect to the given polarization, the $(0,1)$-part of the universal connection $u$ induces a holomorphic structure on $\mathcal{E}_{\C}$.
We refer to $\mathcal{E}_{\C}\to \Sigma \times \mathcal{A}$ as the universal $G_{\C}$-bundle which is furthermore $\Sigma G_{\C}$-equivariant.

\begin{proposition}\label{moment map of universal connection is Id}
The moment map $\mu^{u}$ of the universal connection $u$
is the identity, \ie it maps $\xi\in \Omega^0(\Sigma,\g)$ to $\xi\in \Omega^0(\Sigma,\g)\subset \Omega^0(\Sigma\times \mathcal{A},\g)$ itself.
\end{proposition}

\begin{proof}
The universal connection $u:T\mathcal{E}\to \g$ is the orthogonal projection to the vertical part. 
Given $\xi\in \Omega^0(\Sigma,\g)$, it induces a vertical vector field $\xi_{\mathcal{E}}$ on $\mathcal{E}=Q\times \mathcal{A}$.
 Let $\{e_i\}$ be the standard orthogonal basis of Lie algebra $\g$ under
 the $\operatorname{Ad}(G)$-invariant inner product. Then
 $e_i$ induces the vertical vector field $e_{i,\mathcal{E}}$ on $\mathcal{E}$ with unit norm. We have
\begin{align*}
\mu^{u}(\xi)=u(\xi_{\mathcal{E}})=\sum_i g_{\mathcal{E}}(\xi_{\mathcal{E}},e_{i,{\mathcal{E}}})\cdot e_i.
\end{align*}
Since $e_{i,\mathcal{E}}$ is vertical, evaluated at $(x,A)\in \Sigma \times \mathcal{A}$,
\begin{align*}
g_{\mathcal{E}}(\xi_{\mathcal{E}},e_{i,\mathcal{E}})|_{(x,A)}=\langle \xi(x),e_i \rangle.
\end{align*}
Finally we have
\begin{align*}
\mu^{u}(\xi)|_{(x,A)}=\sum_i \langle \xi(x),e_i\rangle e_i=\xi(x).   
\end{align*}
The moment map $\mu^{u}:\Omega^0(\Sigma,\g) \to \Omega^0(\Sigma\times \mathcal{A},\g)$ is the identity.
\end{proof}

We now consider the curvature $R^u$ of the universal connection $u$, which is a $\g$-valued $\Sigma G$-invariant $2$-form on $\mathcal{E}=Q\times \mathcal{A}$. The space $\Sigma \times \mathcal{A}$ carries a natural almost complex structure—induced by the complex structure on the Riemann surface $\Sigma$ and the Hodge star operator on $T\mathcal{A} \cong \Omega^1(\Sigma,\mathfrak{g})$, which yields a decomposition of differential forms into types $(2,0)$, $(1,1)$, and $(0,2)$. In particular, for $A \in \mathcal{A}$, we have $(R^u)^{2,0}_A=F_A$.

Given a unitary representation $V$, we can define the associated bundle of the universal bundle $(\mathcal{E},u)$, which is denoted by $(\mathcal{E}_V,\nabla_V)\to \Sigma \times \mathcal{A}$.
\begin{center}
\begin{tikzcd}
   & (\mathcal{E}_V,\nabla_V) \arrow[d] \\
 \Sigma\arrow[r]  & \Sigma \times \mathcal{A}\arrow[d] \\
   &    \mathcal{A}
\end{tikzcd}
 $ \rightsquigarrow$
\begin{tikzcd}
   (\mathrm{DET}(V),g^{\mathrm{DET}(V)},\nabla^{\mathrm{DET}(V)})  \arrow[d] \\
   \mathcal{A}
\end{tikzcd}
\end{center}

\begin{theorem}

The equivariant curvature of the Bismut-Freed connection is $R^{\mathrm{DET}(V)}-\sqrt{-1}\mu^{\mathrm{DET}(V)}=d_V(\omega-\mu)$. 

\end{theorem}

\begin{proof}

We apply the equivariant family index theorem \cite{Freed:2016mpb},
    \begin{align*}
  R^{\mathrm{DET}(V)}-\sqrt{-1}\mu^{\mathrm{DET}(V)}&=2\pi\sqrt{-1}\int_{\Sigma}-\frac{1}{8\pi^2}d_V\langle R^u+\mu^u,R^u+\mu^u\rangle \\
    &=d_V(\omega-\frac{\sqrt{-1}}{2\pi}\int_{\Sigma}\langle R^u\wedge \mu^u\rangle). \nonumber 
    \end{align*}
Since $(R^u)_A^{2,0}=F_A$, and Proposition \ref{moment map of universal connection is Id} that $\mu^{u}$ is the identity,
    \begin{align*}
    \mu^V(\xi)(A)= \frac{1}{2\pi}\int_{\Sigma}d_V\langle F_A \wedge \xi\rangle.
    \end{align*}
Then
    \begin{align*}
     R^{\mathrm{DET}(V)}-\sqrt{-1}\mu^{\mathrm{DET}(V)}=d_V(\omega-\mu).
    \end{align*}    
    
\end{proof}

\begin{corollary}\label{moment map eq}

$(\mathrm{DET}(V),g^{\mathrm{DET}(V)},\nabla^{\mathrm{DET}(V)}) $ has the same $\Sigma G$-equivariant curvature with Chern-Simons line bundle $(\mathcal{L},h,\nabla)^{d_V}$. 

\end{corollary}

\subsubsection{Isometry to Chern-Simons line bundle}

\begin{lemma}\label{exist section}

If $(\mathcal{L},h,\nabla)$ is a $\Sigma G$-equivariant line bundle over $\mathcal{A}$, whose moment map and curvature vanish, then 
there exists a trivialization, a $\Sigma G$-equivariant section $s$ of unit norm, which is unique up to a constant phase factor.

\end{lemma}

\begin{proof}

Given $A_0 \in \mathcal{A}$, we fix the value $s(A_0)$. To define the value of $s(A)$ for $A\in \mathcal{A}$, we consider the parallel transport of the flat connection $\nabla$ along any path $\gamma$ from $A_0$ to $A$.

We define $s(A)=P_{\gamma}s(A_0)$ where $P_{\gamma}$ is the parallel transport operator along $\gamma$. In order to prove it is well-defined, we need to show that the value of $s(A)$ is independent of the choice of path $\gamma$.
Since $\mathcal{A}$ is affine, it is simply-connected. Therefore, given $\gamma_0$ and $\gamma_1$ from $A_0$ to $A$, we can find a deformation $\gamma_t$ connecting $\gamma_0$ and $\gamma_1$. Since $\nabla$ is flat, we have $\frac{d}{dt}P_{\gamma_t}=0$. Thus  $P_{\gamma_1}s(A_0)=P_{\gamma_0}s(A_0)$. We define $s(A)$ by parallel transport, thus $s$ is of unit norm and $\nabla s=0$.

The vanishing of the moment map implies, via Kostant’s formula, that the infinitesimal $\Sigma G$-action on $\mathcal{L}$ satisfies $\xi_{\mathcal{L}}(s) = 0$ for all $\xi \in \Omega^0(\Sigma, \mathfrak{g})$, where $\xi_{\mathcal{L}}$ is the fundamental vector field on $\mathcal{L}$ induced by $\xi$.
 Since $\Sigma G$ is connected, $s$ is $\Sigma G$-equivariant. Hence $s$ is unique up to a constant phase
factor which depends on the value of the basepoint $A_0$.

\end{proof}

\begin{proposition}

As $\Sigma G$-equivariant prequantum line bundles over $\mathcal{A}$, we have
\begin{align*}
   (\mathrm{DET}(V),g^{\mathrm{DET}(V)},\nabla^{\mathrm{DET}(V)}) \cong (\mathcal{L},h,\nabla)^{d_V}.
\end{align*}

If $V$ is a real representation, then 
\begin{align*}
(\mathrm{PF}(V),g^{\mathrm{PF}(V)},\nabla^{\mathrm{PF}(V)})\cong (\mathcal{L},h,\nabla)^{d_V/2}.
\end{align*}

\end{proposition}

\begin{proof}

From \eqref{moment map eq}, they share the same equivariant curvature. The isometry follows from the lemma (\ref{exist section}).

\end{proof}

In the genus one case, via the $\Sigma G$-equivariant isomorphism 
\[
(\mathrm{DET}(V), g^{\mathrm{DET}(V)}, \nabla^{\mathrm{DET}(V)}) \cong (\mathcal{L}, h, \nabla)^{d_V},
\]
we can identify the canonical section $\dt_V$ of the determinant line bundle $\mathrm{DET}(V)$ with a section of Chern-Simons line bundle $\mathcal{L}^{d_V}$. Thus we may regard $\mathrm{det}_V$ as an element of $H_{\Sigma G}^0(\mathcal{A},\mathcal{L}^{d_V})$, well-defined up to multiplication by a constant phase factor arising from the ambiguity in the choice of isomorphism. Nevertheless, 
its norm is always well-defined and satisfies
\begin{align*}
   |\mathrm{det}_V(A)|^2=\mathrm{det}_{\zeta}(\slashed\partial_A^{V,*}\slashed\partial^V_A). 
\end{align*} 

\subsection{Push-down Geometry and Genus One Conformal Blocks}\label{Global Model and Conformal blocks}

We now specialize to the elliptic curve $E=\Sigma_{\tau}=\C/(\Z\oplus \tau \Z)$.
What is special about genus one is that the moduli of semi-stable holomorphic principal $G_{\C}$-bundles over an elliptic curve has a global quotient model
\begin{align*}
 M_G[\tau]=(\Sigma_{\tau}\otimes \bigwedge)/W,
\end{align*}
where $\bigwedge=\{t\in \mathfrak{t}|\exp{(2\pi\sqrt{-1}t)}=e \}$ is the co-root lattice of $\g$ and
$W=N_G(T)/T$ is the Weyl group.

In this section, we establish two key facts:
\begin{itemize}

  \item The global quotient model $M_G[\tau] = (\Sigma_\tau \otimes \bigwedge)/W$ can be realized as the quotient of the subspace $\mathcal{C} \subset \mathcal{A}$ of constant Cartan connections.
  
  \item The Looijenga line bundle \cite{Looijenga1976RootSA} on the abelian variety $\Sigma_\tau \otimes \bigwedge$ is naturally isomorphic to the restriction of the Chern-Simons line bundle $\mathcal{L}$.

\end{itemize}

Let $\mathcal{C}=\{\frac{\pi}{\tau_2}wd\bar{z}|w\in \mathfrak{t}_{\C} \}\subset \mathcal{A}$ be the subspace of constant Cartan connections.

Clearly, we have $\mathcal{C}\subset \mu^{-1}(0)$. Moreover, any flat connection is $\Sigma G$-equivalent to a constant Cartan connection. By the Riemann–Hilbert correspondence, flat $G$-connections correspond to representations of the fundamental group $\Hom(\pi_1(\Sigma),G)$; for an elliptic curve, these are determined by pairs of commuting elements in $G$.
Since any pair of commuting elements in $G$ can be simultaneously conjugated into the maximal torus $T$ (cf.\ \cite[Lemma 5.11]{Friedman1998HolomorphicPB}), we obtain the following proposition.

\begin{proposition}\label{Conj into constant connections-}

Any flat connection is gauge-equivalent, under the action of $\Sigma G$, to a constant connection in $\mathcal{C}$.

\end{proposition}

Let $K_{\mathcal{C}}=\{g\in \Sigma G_{\C}|g\mathcal{C}\subset \mathcal{C}\}$
be the stabilizer of $\mathcal{C}$. We show that $K_{\mathcal{C}}$-action on $\mathcal{C}$ is equivalent to the affine Weyl group $W_{\mathrm{aff}}=(\bigwedge\oplus \tau \bigwedge)\rtimes W$-action on $\mathfrak{t}_{\C}$.

Given $\lambda=\lambda_1-\tau \lambda_2\in \bigwedge\oplus \tau\bigwedge$, we define
\begin{align*}
    g_{\lambda}:\C \to G_{\C},\; z\mapsto \exp(\frac{\pi}{\tau_2}[z\bar{\lambda}-\bar{z}\lambda]).
\end{align*}
 
A direct computation shows that $g_{\lambda}(z+1)=g_{\lambda}(z),g_{\lambda}(z+\tau)=g_{\lambda}(z)$. Thus $g_{\lambda}\in \Sigma G_{\C}$. 

Given $\frac{\pi}{\tau_2}wd\bar{z}$ with $w\in \mathfrak{t}_{\C}$, the gauge transformation $g_{\lambda}\in K_{\mathcal{C}}$ acts by the translation
\begin{align*}
    g_{\lambda}(\frac{\pi}{\tau_2}wd\bar{z})=\frac{\pi}{\tau_2}(w+\lambda)d\bar{z}.
\end{align*}

The group $K_{\mathcal{C}}$ is generated by the constant Weyl group elements (via $W \hookrightarrow \Sigma G_{\mathbb{C}}$) and the transformations $\{ g_\lambda \mid \lambda \in \bigwedge \oplus \tau \bigwedge \}$. Under the identification $\mathcal{C}\cong \mathfrak{t}_{\C}$, each $g_{\lambda}$ acts by translation $w\mapsto w+\lambda$ for $w\in \mathfrak{t}_{\C}$, while rhe Weyl group $W$ acts linearly on $\mathfrak{t}_{\mathbb{C}}$ via its natural action on $\mathfrak{t}$, extended $\mathbb{C}$-linearly. 
Hence the $K_{\mathcal{C}}$–action on $\mathcal{C}$ coincides with the affine Weyl group action $W_{\mathrm{aff}}=(\bigwedge\oplus \tau\bigwedge)\rtimes W$ on $\mathfrak{t}_{\C}$. 

Consequently, we obtain an isomorphism of quotient spaces:
\begin{align*}
\mathcal{C}/K_{\mathcal{C}}=\mathfrak{t}_{\C}/W_{\mathrm{aff}}=
    (\Sigma_{\tau}\otimes_{\Z} \bigwedge)/W. 
\end{align*}

Proposition \ref{Conj into constant connections-} and the next proposition show that $\mathcal{C}/K_\mathcal{C}=\mca^{ps}/\Sigma G_{\C}=\mu^{-1}(0)/\Sigma G$ (as sets).

\begin{proposition}

If $A, A'\in \mathcal{C}$ and there exists $g\in \Sigma G_{\C}$ such that $A'=gA$, then there exists $g'\in K_\mathcal{C}$ such that $A'=g'A$.

\end{proposition}

\begin{proof}

Suppose $A'=gA$, then the stabilizer satisfies $K_{A'}=gK_Ag^{-1}$. Since $T\subset K_A$, it follows that $gTg^{-1}\subset K_{A'}$. On the other hand, since $T\subset K_{A'}$, $T$ and $gTg^{-1}$ are both maximal tori of $K_A'$, they must be conjugate to each other, \ie there exists $k'\in K_{A'}$ such that
\begin{align*}
    gTg^{-1}=k'^{-1}Tk.
\end{align*}
Set $g'=k'g$,
then we have
$
g'A=k'gA=k'A'=A'.
$

The next step is to show that $g'\in K_{\mathcal{C}}$.

Since $g'Tg'^{-1}=T$,  we have $g'=g''\cdot w$ where $w\in N_G(T)$ and $g''\in \Sigma T_{\C}$.
Set $A''=wA$. Then $A''$ is still in $\mathcal{C}$.
We only need to prove a weaker statement
that if there is $g''\in \Sigma T_{\C}$ such that $A''=g''A'$ where $A',A''\in \mathcal{C}$, then $g''\in K_{\mathcal{C}}$.

In this case, for $A'(A'')=\frac{\pi}{\tau_2}a'(a'')d\bar{z}$ where $a'(a'')\in \mathfrak{t}_{\C}$,
from
\begin{align*}
A''=g''A'=A'-\bar{\partial}g''(g'')^{-1},
\end{align*}
we obtain that 
\begin{align*}
    \frac{\pi}{\tau_2}(a'-a'')=\partial_{\bar{z}}g'' (g'')^{-1}\in \mathfrak{t}_{\C}.
\end{align*}
This equation has no solution has no solution unless $a'-a''\in \bigwedge\oplus \tau \bigwedge$.
When $a'-a''=\lambda=\lambda_1-\tau \lambda_2\in \bigwedge\oplus \tau \bigwedge$,
$g''$ is solved by $g''(z)=g_{\lambda}(z)=\exp(\frac{\pi}{\tau_2}[z\bar{\lambda}-\bar{z}\lambda])$ which lies in $\Sigma T_{\C}$.
Therefore $g'=k'g\in K_{\mathcal{C}}$, as required.

\end{proof}

The genus one conformal block $V_k(G)$ can be described as $K_{\mathcal{C}}$-equivariant holomorphic sections of Chern-Simons line bundle restricted to the space of constant Cartan connections $\mathcal{C}$. 
Let $\{e_i\},i=1\cdots r$ be the co-root basis of $\mathfrak{t}$, and write 
$z=z^ie_i$. The map $z\mapsto \{z^i\}$ gives the identification between $\mathfrak{t}_{\C}$ and $\C^{r}$. We always identify $\mathcal{C}$ with $\C^r$. 
Let $C_{ij}=\langle e_i,e_j\rangle$ be the normalized Cartan matrix  and $C^{ij}$ be its inverse. 
The symplectic structure on $\mathfrak{t}_{\C}$ is given by the restriction of $\omega$ to the subspace of constant $\mathfrak{t}_{\C}$-valued connections. The corresponding symplectic structure $\omega_0$ on $\C^r$ is given by
$
\omega_0=\frac{i\pi}{\tau_2}C_{ij}dz^id\bar{z}^j.
$

Under the identification $\mathcal{C}\cong \C^r$, the Chern-Simons line bundle is denoted by $\widetilde{\mathcal{L}_0}$, which is still a prequantum-line bundle and the holomorphic structure is determined by the polarization given by the $(1,0)$-part of the connection.
The action of the affine Weyl group $W_{\mathrm{aff}}=(\bigwedge\oplus \tau \bigwedge)\rtimes W$ on $\C^r$ can be lifted to $\widetilde{\mathcal{L}_0}$.
Denote by $\mathcal{L}_0$ the descent of $\widetilde{\mathcal{L}_0}$. This is a basic line bundle on the abelian variety $\Sigma_{\tau}\otimes_{\Z} \bigwedge$, as in \cite{Looijenga1976RootSA}.

We have the following:
\begin{align*}
H^0_{K_{\mathcal{C}}}(\mathcal{C},\mathcal{L}^k)=
H_{W_{\mathrm{aff}}}^0(\mathfrak{t}_{\C},\widetilde{\mathcal{L}_0}^k)=H^0(\Sigma_{\tau}\otimes \bigwedge,\mathcal{L}_0^k)^W.
\end{align*}
Since the stabilizer $K_A$ acts trivially on the fiber of $\mathcal{L}_0$ at $A$, the line bundle $\mathcal{L}_0$ descends to a line bundle $\widetilde{\mathcal{L}}$ on $M_G[\tau]=(\Sigma_{\tau}\otimes \bigwedge)/W$.

\begin{definition}\label{Conformal Block}
The genus one conformal block $V_k(G)$ is defined as 
\begin{align}
V_k(G):=H^0(M_G[\tau],\widetilde{\mathcal{L}}^k)=H^0(\Sigma_{\tau}\otimes \bigwedge,\mathcal{L}_0^k)^W.
\end{align}
\end{definition}

The genus one case is rather special, as both the coarse moduli $M_G[\tau]$ and the conformal block $V_k(G)$ have explicit descriptions. Specifically, $V_k(G)$ is spanned by Kac-Weyl characters of level $k$ integrable highest weight representations of the affine Lie algebra associated to $\g$.
 We will recall this fact in Subsection \ref{subsec:q-conj class} from the perspective of representation theory.

Moreover, the coarse moduli $M_G[\tau]=(\Sigma_{\tau}\otimes \Lambda)/W$ is a weighted projective space whose weights are determined by the root system of $\g$. This is a theorem due to Looijenga and Bernshtein-Schwartsman(cf.\cite[Theorem 4.3]{Friedman1997PrincipalGB}). From the weight data, $M_G[\tau]=(\Sigma_{\tau}\otimes \Lambda)/W$ is simply connected and its Picard group is $\Z$.

\section{Geometric Invariant Theory by Yang-Mills Flow}\label{GIT of Yangmills flow}

In this section, we review some basic properties of Yang-Mills flow.

\subsection{Basics of Yang-Mills Flow}

\begin{definition}

The Yang-Mills functional is the squared norm of the moment map $\mu$:
\begin{align*}
\mathrm{YM}:\mathcal{A}\to \mathbb{R},\; A\mapsto \int_\Sigma\langle F_A\wedge*F_A\rangle.
\end{align*}

The Yang-Mills flow is defined as the negative gradient flow of Yang-Mills functional:
\begin{align*}
\frac{dA(t)}{dt}=-\nabla \YM({A(t)}).
\end{align*}

\end{definition}

The gradient of the Yang-Mills functional satisfies:
\begin{align*}
\nabla\YM(A)=d_A^*F_A=*d_A*F_A.
\end{align*}
 
\begin{definition}
 Let $A_0 \in \mathcal{A}$. We call $A(t) \in C^0([0, \infty), \mathcal{A}) $ a weak solution of the initial value problem
\begin{align}\label{initial problem}
\frac{dA(t)}{dt}=-\nabla \YM({A(t)})= -*d_{A(t)}*F_{A(t)}, \quad A(0) = A_0.
\end{align}
if
 there exists a sequence $A_k$ of smooth solutions which converges in $C_{loc}^0([0, \infty), \mathcal{A}) $ to $A$, \ie for every closed interval $[0,T]$, we have $$\lim\limits_{k\to \infty}\sup\limits_{t\in [0,T]}\vert| A_k(t)-A(t)\vert|_{W^{1,2}}=0.$$
\end{definition}

Johan Rade proved the global existence of the Yang-Mills flow and its convergence on Riemann surfaces \cite{Rade1992OnTY}. In addition, if the initial data is smooth, then the flow is also smooth.

\begin{theorem}[{\cite{Rade1992OnTY}}]\label{Yang-Mills Flow} 

There exists a unique weak solution $A(t) \in C_{\text{loc}}^0([0, \infty), \mathcal{A})$ for the initial value problem \eqref{initial problem}.
The solution converges in $W^{1,2}$-topology to an element $A_\infty\in \mca$. This flow satisfies the Lojasiewicz type estimate: there exists some constant $C, \beta>0$ such that
	\begin{align*}
	||A(t)-A_\infty||_{W^{1,2}}\leqslant Ct^{-\beta}.  
	\end{align*}
 Furthermore, if $A_0$ is smooth, then the solution $A(t)$ is smooth and depends smoothly on the initial data $A_0$. Also $A_\infty\in \mca$.
 
	\end{theorem}
    
	Using the Yang-Mills flow, we can prove some theorems analogous to the basic results in finite-dimensional GIT theory.

	Here are some basic results in infinite-dimensional GIT that will be used later. For details, we refer to the survey \cite[Theorem 4.14, 4.15]{Trautwein2017ASO}.
\begin{theorem}[Moment Limit Theorem]\label{Moment Limit Theorem}
Suppose $A_0\in \mathcal{A}$ and let $A(t) \in C_{\text{loc}}^0([0, \infty), \mathcal{A})$ be the solution of the Yang-Mills flow starting at $A_0$. The limit of the Yang-Mills flow satisfies
\begin{align*}
\YM(A_\infty)=\inf\limits_{g\in \Sigma G_\mbc} \YM(g A_0).
\end{align*}
\end{theorem}

\begin{theorem}[Ness Uniqueness Theorem]\label{Ness Uniqueness Theorem}
Suppose $A_0\in \mca$ and $A_1, A_2$ are contained in the $W^{1,2}$-closure of $\Sigma G_\mbc A_0$ satisfying
\begin{align*}
\YM(A_1)=\YM(A_2)=\inf\limits_{g\in \Sigma G_\mbc} \YM(g A_0),
\end{align*}        
then $A_1, A_2$ are on the same $\Sigma G_{\C}$-orbit.
\end{theorem}

We can define the notion of stability using Yang-Mills flow.

    \begin{definition}\label{stability by Yang-Mills}
        Let $A_\infty$ be the limit of Yang-Mills flow starting from $A_0$. Then we call $A_0$
        \begin{itemize}
            \item semi-stable if $\YM(A_\infty)=0$;
            \item poly-stable if $A_0$ is semi-stable and $A_\infty\in \Sigma G_\mbc A_0$;
            \item stable if $A_0$ is poly-stable and the stabilizer of $\Sigma G_\mbc$ at $A_0$ is discrete;
            \item unstable if $A_0$ is not semi-stable.
        \end{itemize}
    \end{definition}

\subsection{Connection/Bundle Correspondence}

We define the stability of a holomorphic principal $G_{\C}$-bundle by the slope stability of its adjoint bundle.

\begin{definition}[{\cite[Definition 2.1]{Friedman1998HolomorphicPB}}]\label{Slope stability}

We call a holomorphic principal $G_{\C}$-bundle $P\to \Sigma$ stable (semi-stable, poly-stable, unstable) if its associated adjoint bundle $\mathrm{ad}P$ is stable (semi-stable, poly-stable, unstable).

\end{definition}

When $G$ is semisimple, the following theorem asserts that the stability of any associated bundle can be tested solely by the adjoint bundle.

\begin{theorem}[{\cite[Theorem 2.3]{Friedman1998HolomorphicPB}}]

Let $P$ be a holomorphic $G_{\C}$-bundle. Then the following are equivalent:

\begin{itemize}

\item $P$ is semi-stable.

\item For every finite-dimensional vector space $V$ and representation $\phi:G\to GL(V)$, the associated bundle $P(V)$ is semi-stable.

\end{itemize}
   
\end{theorem}

The following generalized Narasimhan-Seshadri-Ramanathan theorem shows that the stability of the connection $A$ defined by the convergence of Yang-Mills flow in Definition \ref{stability by Yang-Mills} is equivalent to the stability of the corresponding principal $G_\mbc$-bundle in Definition \ref{Slope stability}.
    
\begin{theorem}[{cf.\cite[Theorem 3.10]{Trautwein2017ASO}}]

The connection $A$ is stable (semi-stable, poly-stable, unstable) if and only if the corresponding $G_{\C}$-bundle $P_A$ is stable (semi-stable, poly-stable, unstable).

\end{theorem}

Let $\mathcal{A}^{ss}$ be the set of semi-stable points in $\mathcal{A}$ and $\mathcal{A}^{ps}$ be the set of poly-stable points in $\mathcal{A}$. Then $\mathcal{A}^{ps}$ equals $\Sigma G_{\C}\cdot \mu^{-1}(0)$, the $\Sigma G_{\C}$-orbits of flat connections.

\begin{definition}

The homomorphism $\rho\in Hom(\pi_1(\Sigma),G)$ defines a flat $G$-bundle $P_{\rho}$ over $\Sigma$ by the Riemann-Hilbert correspondence. Using the inclusion $G\subset G_{\C}$, we can view $P_{\rho}$ as a holomorphic $G_{\C}$-bundle. Bundles from this construction are called flat bundles.

\end{definition}

From the definition, any poly-stable connection is $\Sigma G_{\C}$-equivalent to a flat connection and the Riemann-Hilbert correspondence shows that $A\in \mathcal{A}^{ps}$ if and only if the corresponding $G_{\C}$-bundle $P_A$ is flat.  

We recall the definition of S-equivalence. 

A family of holomorphic principal $G_{\C}$-bundles over $\Sigma$ parametrized by a complex space $X$ is a $G_{\C}$-bundle $\mathcal{P}$ over $\Sigma\times X$. 
Based on the construction of the universal bundle in (\ref{universal bundles}), the holomorphic map $f:X\to \mathcal{A}$ defines such a family as holomorphic principal $G_{\C}$-bundle $\mathcal{P}_f\to \Sigma \times X$, where $\mathcal{P}_f\to \Sigma\times X$ is the pull-back of the universal $G_{\C}$-bundle $(\mathrm{Id}\times f)^*\mathcal{E}_{\C}$.

Under the GIT-quotient, the equivalence of two connections $A_1,A_2\in \mathcal{A}^{ss}$ is,
let $\mathcal{O}_1,\mathcal{O}_2$ be the corresponding $\Sigma{G_\C}$ orbits of $A_1,A_2$,
$A_1\sim A_2$ if $\overline{\mathcal{O}_1}\cap \overline{\mathcal{O}_2}\neq \emptyset$. 
The equivalence of connections coincides with the S-equivalence of $G_{\C}$-bundles in \cite{Ramanathan1975StablePB}, \cite[Defintion 1.8]{Friedman1998HolomorphicPB}.
We say that the semi-stable $G_{\C}$-bundles $P$ and $P'$ are S-equivalent if there exists a family of semi-stable $G_{\C}$-bundles $\mathcal{P}$ parametrized by an irreducible complex space $X$ and a point $s\in X$ such that $\mathcal{P}_t\cong P$ for all $t\neq s$, while $\mathcal{P}_s\cong P'$.

\begin{table}[h!]
    \centering
    \begin{tabular}{|c|c|c|c|c|}
    \hline
                  & Semi-stable & Poly-stable & Equivalence in GIT-quotient \\ \hline
    $A\in \mathcal{A}$ & $\mathrm{YM}(A_{\infty})=0$& $A\in \Sigma G_{\C}\mu^{-1}(0)$ & Closures of orbits intersect \\ \hline
    $G_{\C}$-bundle $P_A$ &$P_A$ is semi-stable & Flat & S-equivalence by degeneration \\ \hline
    \end{tabular}
    \caption{S-equivalence and GIT-quotient}
    \label{tab:3x4}
 \end{table}

It follows that the natural map from the set of S-equivalence classes of semi-stable $G_{\C}$-bundles to $M_G[\tau]=\mu^{-1}(0)/\Sigma G$ is bijective. Furthermore, as a direct corollary of Theorem \ref{Yang-Mills Flow}, the natural projection defined by the S-equivalence (or Yang-Mills flow equivalently) $\mathcal{S}: \mca^{ss}\to M_G[\tau]$ is continuous with respect to the $W^{1.2}$ topology. In fact, $\mathcal{S}$ is holomorphic, which implies that $M_G[\tau]$ is the moduli space of S-equivalence classes of semi-stable $G_\mbc$-bundles.

\section{Proof of Quantization Commutes with Reduction}\label{Sec:Proof}

In this section, we mainly prove $[Q,R]=0$ in the genus one case. Namely, there is a natural isomorphism of $\C$-vector spaces \[\mathrm{r}: \mathcal{H}_k(G)\to V_k(G).\]

\subsection{Equivariant Holomorphic Sections along Yang-Mills Flow}

We show that the norm of equivariant holomorphic sections of the Chern-Simons line bundle is non-decreasing along the Yang-Mills flow. This is the infinite-dimensional analogy of \cite{SjamarrSlice},

\begin{lemma}\label{IncreasingNorm}

Suppose $A(t)$ is a Yang-Mills flow starting from $A(0)=A_0\in\mca$ and $s\in H^0_{\Sigma G}(\mathcal{A},\mathcal{L}^k)$, then $|s_{A(t)}|\geqslant|s_{A(t')}|$ if $t>t'$.

\end{lemma}

\begin{proof}

For any $\xi\in \Sigma\g$, the $\Sigma G_{\C}$-invariance of $s$ implies that $(i\xi)^{\#}\cdot s=0$, and hence $\nabla_{*\xi^{\#}}s=-\langle \mu,\xi\rangle s$. It follows that
\begin{align}\label{Derivative}
(i\xi)^{\#}\abs{s}^2=2\langle\nabla_{*\xi^{\#}} s,s \rangle=-2\langle \mu,\xi\rangle\abs{s}^2.
\end{align}
Along the flow, since $\frac{dA(t)}{dt}=-*d_{A(t)}*F_{A(t)}=-(i*F_{A(t)})^{\#}$, we have
		\begin{align*}
		    \frac{d|s(A(t))|^2}{dt}=-(i*F_{A(t)})^{\#}|s|^2.
		\end{align*}
        Take $\xi=*F_{A(t)}$ in \eqref{Derivative}, we get
        \begin{align*}
             \frac{d|s(A(t))|^2}{dt}=2\langle \mu,*F_{A(t)}\rangle\abs{s}^2=\frac{|s|^2}{\pi}\int_{\Sigma}\Tr (F_{A(t)}\wedge *F_{A(t)})=\frac{\YM(A(t))}{\pi}|s|^2
        \geqslant 0.  
        \end{align*}     
        
\end{proof}

An immediate corollary is that any equivariant holomorphic section must vanish on the unstable locus:

\begin{corollary}\label{Vanishing on us}

For any $s\in H^0_{\Sigma G}(\mca,\mcl^k)$ and $A_0$ in $\mca^{us}$, we have $s(A_0)=0$.

\end{corollary}
    
\begin{proof}

Consider the Yang-Mills flow $A(t)$ with $A(0)=A_0$. The limit $A_\infty$ is a critical point of the Yang-Mills functional, so $(i*F_{A_\infty})^{\#}=0$. Thus, we have
\begin{align*}
0=(i*F_{A_\infty})^{\#}\langle s,s \rangle=-2\YM(A_\infty)|s(A_\infty)|^2/\pi.
\end{align*}
But by Theorem \ref{Moment Limit Theorem} and $A_0\in \mca^{us}$, we have $\YM(A_\infty)\neq 0$, and hence $s(A_\infty)=0$. Now, by Lemma \ref{IncreasingNorm}, $|s(A_0)|\leqslant |s(A_\infty)|$, which implies $s(A_0)=0$.

\end{proof}

Later, in Proposition \ref{regularized det increases}, we will apply this result to the $\zeta$-regularized determinant of the Laplacian. 
A further corollary is that any equivariant holomorphic section is actually determined by its values on the space of constant Cartan connections $\mathcal{C}$.

\begin{corollary}\label{Injectivity}

For any $s,s'\in H^0_{\Sigma G}(\mca,\mcl^k)$, if $s|_\mathcal{C}=s'|_\mathcal{C}$, then $s=s'$.

\end{corollary}

\begin{proof}
    
Let $s''=s-s'$, then $s''\in H^0_{\Sigma G}(\mca,\mcl^k)$. On $\YM^{-1}(0)=\Sigma G\cdot \mathcal{C}$, the norm squared satisfies $|s''|^2=0$ since the $\Sigma G$-action does not change the norm.

If $A_0\in\mca^{ss}$, consider the Yang-Mills flow $A(t)$ with $A(0)=A_0$. By the definition of semi-stable connections, we have $A_\infty\in\YM^{-1}(0)$. Then, by Lemma \ref{IncreasingNorm}, $|s''(A_0)|\leqslant |s''(A_\infty)|=0$. On the other hand, by Lemma \ref{Vanishing on us}, $s''(A)=0$ for all $A\in\mca^{us}$. Thus $s=s'$ on $\mca$.

\end{proof}

\subsection{Regular and Poly-Stable Locus}

We first recall that, for a finite-dimensional complex manifold, the complex analytic subsets are precisely the closed subsets of the (analytic) Zariski topology.

\begin{definition}\label{Zariski open}

A subset $U$ of $\mca$ is called Zariski open if for any holomorphic map $f: X\to \mca$ from a connected finite dimensional complex manifold $X$, $f^{-1}(U)$ is an (analytic) Zariski open subset of $X$.

\end{definition}

Similar to finite dimensional GIT, the semi-stable locus is Zariski open in this definition. But we only need the following weaker result for our proof:

\begin{lemma}\label{SS Openness}

The semi-stable $\mca^{ss}$ contains a Zariski open subset.

\end{lemma}

\begin{proof}

By Corollary \ref{Vanishing on us}, if there exists $s\in H^0_{\Sigma G}(\mca,\mcl^k)\neq 0$ for some $k$, then $\mca^{us}$ is contained in the zero locus of $s$, which is a Zariski closed subset of $\mca$.

Recall that in Theorem \ref{detpf}, the canonical sections of the determinant line bundle under the isometry to the specific level of the Chern-Simons line bundle have been constructed. These are equivariant holomorphic sections, and the lemma follows.

\end{proof}

For higher genus, the space of stable connections is Zariski open by \cite{Ramanathan1975StablePB}. 
However, in the genus one case, there is no stable connection. The space of regular and poly-stable connections is the suitable replacement.

Now, we introduce the notion of regular connections.

The Lie algebra $k_A$ of the stabilizer $K_A=\{g\in \Sigma G_{\C},gA=A\}$ is identified with 
\begin{align*}
k_A\cong H^0(\Sigma,V_A),
\end{align*}
where $V_A$ is the adjoint bundle of the principal $G_\mbc$-bundle $P_A$ corresponding to $A$.

Recall that a connection $A$ is stable if $A$ is poly-stable and the stabilizer group $K_A$ is discrete. This is equivalent to saying that $A$ is poly-stable and the dimension $H^0(\Sigma,V_A)$  attains its minimum value, which is zero.

In the genus one case, the stable part $\mca^s$ is empty, since $H^0(\Sigma,V_A)\geqslant r$ when $A\in \mca^{ss}$, where $r$ is the rank of $G$.

Therefore, it is natural to consider the connections whose stabilizers have the smallest dimension.

\begin{definition}

We call $A\in \mathcal{A}$ regular if $\mathrm{dim} H^0(E,V_A)$ achieves the minimum value $r$. We denote by $\mathcal{A}^{reg}$ the subset consisting of regular connections and by $\mathcal{A}^{rp}$ the subset consisting of regular and poly-stable connections.

\end{definition}

\begin{lemma}\label{Reg Openness}

$\mca^{reg}\subset \mca$ is Zariski open.

\end{lemma}

\begin{proof}

Given a holomorphic map $f: X\to \mca$, the pull-back of the universal bundle $\mathcal{P}_f=f^*(\mathcal{E}_{\C})$ is the corresponding $G_{\C}$-bundle over $E\times X$. Denote its adjoint bundle by $\mathscr{E}$ over $E\times X$. By the upper-semicontinuity theorem, the minimum value locus $f^{-1}(\mca^{reg})$
is Zariski open in $X$.

\end{proof}

Furthermore, we will show that $\mathcal{A}^{rp}$ also contains a non-empty Zariski open subset of $\mathcal{A}$.

We first describe the splitting principle of poly-stable bundles.

By Proposition \ref{Conj into constant connections-}, poly-stable $G_{\C}$-bundles over elliptic curves are all split, \ie the structure group reduces to the Cartan subgroup which is a maximal complex torus. In other words, when $A\in \mathcal{A}^{ps}$, the adjoint bundle $V_A$ splits into degree zero line bundles.

\begin{proposition}\label{Splitting of poly-stable}

If $A\in \mca^{ps}$, there exists a suitable gauge transformation $g\in \Sigma G_{\C}$ such that $gA\in \mathcal{C}$ is a constant Cartan connection represented by some $\mu \in \mathfrak{t}_{\C}$.
The adjoint bundle $V_A$ admits the splitting
\begin{align*}
V_A=\mathcal{O}_E^r\oplus\bigoplus_{\alpha}L_{\alpha(\mu)},
\end{align*}    
where $\alpha$ runs through the root of $\g$ and $L_{\alpha(\mu)}$ is the corresponding degree zero holomorphic line bundle given by the Abel-Jacobi map.

Moreover, $A$ is regular if and only if $\alpha(\mu) \neq 0$ for every root $\alpha$ of $\mathfrak{g}$.

\end{proposition}

We now describe the semi-stable and regular $G_{\C}$-bundles over elliptic curve $E$.

\begin{definition}

Over the elliptic curve $E$, there is a unique indecomposable holomorphic vector bundle of rank $n$ whose Jordan-H\"older constituents are all isomorphic to $\mso_E$. We denote it by $I_n$. 

\end{definition}

Then $I_n$ has a canonical filtration 
\begin{align*}
    \{0\}\subset F^1\subset F^2 \cdots \subset F^n=I_n,
\end{align*}
where $F^i\cong I_i$ and $F^{i+1}/F^i\cong \mso_E$.

For $A\in \mathcal{A}^{ss}$, let $A_{\infty}\in \mathcal{A}^{ps}$ be the unique poly-stable connection given by the limit of Yang-Mills flow.
Since $A_{\infty}$ is poly-stable, by Proposition \ref{Splitting of poly-stable},
the adjoint bundle has the splitting
\begin{align}\label{Split bundle}
V_{A_{\infty}}=\mso_E^r\oplus\bigoplus_{\alpha}L_{\alpha(\mu)},
\end{align}
The Lie algebra $k_{A_{\infty}}=H^0(E,V_{A_{\infty}})$ of the stabilizer group of $K_{A_{\infty}}$ is then identified with 
\begin{align*}
    \mathfrak{h}\oplus \bigoplus_{\alpha,\alpha(\mu)=0}\g^{\alpha},
\end{align*}
where $\g^{\alpha}$ is the root space corresponding to the root $\alpha$.
The stabilizer group $K_{A_{\infty}}$, up to isogeny, is a product $G_1\times \cdots \times G_N$ where $G_i$ is either a simple Lie group or $\mathbb{C}^*$.
Let $r_i$ be the rank of $G_i$ and $d_{ij},1\leqslant j\leqslant r_i$ be the Casimir weights of $G_i$. Moreover, $d_{i1}=1$ if $G_i=\C^*$.
The semi-stable and regular bundle $V_A$ has the following description:

\begin{lemma}[{\cite[Proposition 3.8]{Friedman1997PrincipalGB}}]\label{Regular Representation}

Let $A \in \mathcal{A}$. Then $A\in \mca^{reg}\cap \mca^{ss}$ if and only if the maximal subbundle of the adjoint bundle $V_A$ that admits a filtration with successive quotients isomorphic to $\mso_E$ is given by
\begin{align}\label{Maximal subbundle}
(V_A)_{\mso_{E}}=\bigoplus_{i=1}^N\bigoplus_{j=1}^{r_i}I_{2d_{ij}-1}.
\end{align}
Furthermore, $A\in \mca^{reg}\cap \mca^{ps}$ if and only if $d_{ij}=1$  for all $i$ and $j$. 

\end{lemma}

\begin{corollary}

If $A\in \mca^{reg}\cap \mca^{ss}$, then the adjoint bundle $V_A$ contains a trivial subbundle of rank $r$.

\end{corollary}

\begin{proof}

By the Jordan-H\"older filtration, $F^1=\mso_E$ is a trivial subbundle of $I_n$.
According to Lemma \ref{Regular Representation}, $(V_A)_{\mso_E}$ contains a subbundle $\mso_E^r$.

\end{proof}

From this description, we deduce that the set of regular and polystable connections 
$\mathcal{A}^{\mathrm{rp}} = \mathcal{A}^{\mathrm{reg}} \cap \mathcal{A}^{\mathrm{ps}}$ contains a Zariski open subset.

\begin{proposition}\label{RP Zariski open}

$\mathcal{A}^{rp}$ contains a non-empty Zariski open set of $\mathcal{A}$. 

\end{proposition}

\begin{proof}

By Lemma \ref{SS Openness} and Lemma \ref{Reg Openness}, $\mca^{ss}\cap\mca^{reg}$ contains a Zariski open subset of $\mca$. It suffices to show that $\mca^{rp}$ is a Zariski open subset of $\mca^{ss}\cap\mca^{reg}$.
             
If $A\in \mca^{ss}\cap\mca^{reg}$ but $A$ is not poly-stable, then at least one $d_{ij}$ in \eqref{Maximal subbundle} is larger than $1$, which implies $\dim H^0(E, V_A/\mso_{E}^r)>0$. 
         On the other hand, if $A\in \mca^{rp}$, then $\dim H^0(E, V_A/\mso_E^r)=0$.

Given any holomorphic map $f: X\to \mca^{ss}\cap\mca^{reg}$, the pull-back of universal bundle $\mathcal{P}_f=f^*(\mathcal{E}_{\C})$ is the corresponding $G_{\C}$-bundle over $E\times X$.
Denote its adjoint bundle over $E\times X$ by $\mathscr{E}$.
Under the projection $\mathrm{pr}:E\times X\to X$,
$\mathscr{E}':=\mathrm{pr}^*\mathrm{pr}_*\mathscr{E}$ is a rank $r$-subbundle of $\mathscr{E}$.
For $x\in X$, the fiber $\mathscr{E}'_{x}\to E$ is isomorphic to $\mso_E^r$.

Thus, $f(x)\in \mca^{rp}$ if and only if $\dim H^0(E, (\mathscr{E}/\mathscr{E}')_{x})=0$. 
By the upper-semicontinuity theorem of higher direct image, $f^{-1}(\mca^{rp})$ is Zariski open in $X$.

\end{proof}

\subsection{ The Proof of Main Theorem}

Finally, we prove the main Theorem \ref{QR=0}
\begin{align*}
        	\mathcal{H}_k(G)=V_k(G).
\end{align*}

\begin{proof}[Proof of Theorem \ref{QR=0}]

By definition \ref{Conformal Block}, the genus one conformal block $V_k(G)$ can be identified with the space of equivariant holomorphic sections defined over the space $\mathcal{C}$ of constant Cartan connections
\begin{align*}
    V_k(G)= H^0_{K_{\mathcal{C}}}(\mathcal{C},\mcl^k).
\end{align*}
Thus, it suffices to show that the restriction map 
\begin{align*}
   \mathrm{r}: H^0_{\Sigma G}(\mca,\mcl^k)\to H^0_{K_{\mathcal{C}}}(\mathcal{C},\mcl^k)
\end{align*} 
is bijective.

Injectivity follows directly from Corollary \ref{Injectivity}. To prove surjectivity, we construct the inverse map of $\mathrm{r}$ by showing that every equivariant holomorphic section in $H^0_{K_{\mathcal{C}}}(\mathcal{C},\mcl^k)$ extends equivariantly and holomorphically to $\mca$. 
The key idea is that boundedness and Zariski-openness imply such an extension. 

Note that $\mathcal{C}^{reg}:=\mathcal{C}\cap\mca^{rp}\cong \{z\in \mathfrak{t}_{\C}, \alpha(z)\neq 0,\forall \text{ root }\alpha\}$ is Zariski open in $\mathcal{C}\cong \mathfrak{t_{\C}}$. Consequently, the restriction map $H^0_{K_{\mathcal{C}}}(\mathcal{C},\mcl^k)\to H^0_{K_{\mathcal{C}}}(\mathcal{C}^{reg},\mcl^k)$ is injective. Any element in $H^0_{K_{\mathcal{C}}}(\mathcal{C}^{reg},\mcl^k)$ can extend uniquely to an equivariant holomorphic section over $\mca^{rp}$ 
since the $K_{\mathcal{C}}$-action is given by the $\Sigma G_{\C}$-action on $\mathcal{L}$. Hence, any $s_0\in H^0_{K_{\mathcal{C}}}(\mathcal{C},\mcl^k)$ gives a section $s\in H^0_{\Sigma G}(\mca^{rp},\mcl^k)$. 

To extend $s$ to an equivariant holomorphic section $s' \in H^0_{\Sigma G}(\mathcal{A}, \mathcal{L}^k)$, we require the following boundedness lemma.

\begin{lemma}\label{Boundedness}

For any $s\in H^0_{\Sigma G}(\mca^{rp},\mcl^k)$ coming from $H^0_{K_{\mathcal{C}}}(\mathcal{C}^{reg},\mcl^k)$, there exists a constant $C>0$ (depending on $s$) such that $\abs{s}^2<C$.

\end{lemma}

\begin{proof}

$s$ is the extension of an equivariant section $s_0\in H^0_{K_{\mathcal{C}}}(\mathcal{C},\mcl^k)$ and $s|_{\mathcal{C}^{reg}}=s_0|_{\mathcal{C}^{reg}}$. 
 $|s_0|^2$ is bounded since it is pulled back from the complete variety $M_G[\tau]$.
Therefore, on the space of flat connections $\YM^{-1}(0)=\Sigma G\cdot \mathcal{C}$, $|s|^2$ is bounded because the $\Sigma G$ action is unitary.
By Theorem \ref{Moment Limit Theorem}, for any $A_0\in\mca^{rp}\sse\mca^{ss}$, the Yang-Mills flow starting from $A_0$ converges to a connection $A_\infty\in\YM^{-1}(0)$. Lemma \ref{IncreasingNorm} implies that $|s(A_\infty)|\geqslant|s(A_0)|$, and the boundedness follows.

\end{proof}

Now consider the collection of all finite-dimensional complex subspaces $V$ of $\mca$ such that $ \mca^{rp}\cap V\neq\emptyset$, and denote it by $\msv$. 
By Theorem \ref{RP Zariski open}, $\mca^{rp}\cap V$ contains a nonempty Zariski open subset of $V$, and $s|_{\mca^{rp}\cap V}$ is a holomorphic section of $\mcl^k|_{\mca^{rp}\cap V}$. By Lemma \ref{Boundedness}, $s$ is bounded on $\mca^{rp}\cap V$. Using Riemann's extension theorem, $s$ extends uniquely to a holomorphic section of $\mcl^k|_V$.

We now patch these extensions together to obtain a global extension $s'\in H^0(\mca,\mcl^k)$:
		Indeed, for any two subspaces $V_1, V_2\in \msv$, we can always find a larger subspace $V_3\in \msv$ that contains $V_1, V_2$. For example, one may take $V_3$ to be the subspace spanned by $V_1$ and $V_2$. By uniqueness of the extension, the extensions of $s$ to $V_1$ and $V_2$ agree on overlaps, since both are restrictions of the extension of $s$ to $V_3$.
		
The section	$s'$ is in fact $\Sigma G$-equivariant: 
    For any $g$ of $\Sigma G$, $gs'$ is the unique extension of $gs$. But $gs=s$, hence $gs'=s'$. Therefore, $s'\in H^0_{\Sigma G}(\mca,\mcl^k)$ is the extension of $s$.
    
We have thus constructed a unique extension $s'$ of $s_0$, and by construction the restriction $\mathrm{r}(s')$ is exactly $s_0$. Consequently, the assignment $s_0 \mapsto s'$ defines the inverse of $\mathrm{r}$, hence $\mathrm{r}$ is bijective.

\end{proof}

\begin{remark}

If $\Sigma$ is a Riemann surface with genus $g(\Sigma)>1$, we can still prove that quantization commutes with reduction by the same method. The only difference is that a global slice $\mathcal{C}$ may no longer exist. Nevertheless, Ramanathan proved that:
    \begin{itemize}
    
        \item The stable locus $\mca^s$ is Zariski open (in the sense of definition \ref{Zariski open}). This is essentially \cite[Proposition 4.1]{Ramanathan1975StablePB}.
        
        \item For any $A\in\mca^s$, there is a local slice given by a holomorphic map from a neighborhood $U$ of $0$ in $H^1(\Sigma, V_A)$ to $\mca$. This is essentially \cite[Theorem 4.2]{Ramanathan1975StablePB}.

        \item There is a holomorphic projection from $\mca^s$ to the moduli space of stable $G_\mbc$-bundles. This is essentially \cite[Theorem 4.3]{Ramanathan1975StablePB}.    
    \end{itemize}
    
    Therefore, one can replace $\mca^{rp}$ in the proof with $\mca^s$, pull every section on the coarse moduli back to $\mca^s$, and then extend it to a global equivariant holomorphic section.

\end{remark}

\section{Further Discussions}\label{Sec:Further Discussions}

\subsection{Regularized Determinants along Yang-Mills Flow}\label{regularized determinants increase}

The construction of the determinant line bundle can be generalized to allow twisting by any degree-zero line bundle, not just the square root of the canonical line bundle $\sqrt{K}$.

Fix a K\"ahler metric on the elliptic curve $E$, and let $L_z$ denote the degree-zero holomorphic line bundle corresponding to $z\in E$ via the Abel–Jacobi map.
 Let $V$ be a finite-dimensional representation of $G$, and let $A\in \mca$ be a connection. 
 Denote by $\bar{\partial}_{A,z}^V$ the $\bar{\partial}$-operator on $V_A \otimes L_z$;
its associated Hodge Laplacian is
\begin{align*}
\Delta_{A,z}^V=(\bar{\partial}_{A,z}^V)^{*}\,\bar{\partial}_{A,z}^V.
\end{align*}

The construction of the determinant line bundle in Theorem \ref{detpf} can be generalized from the spin structure $\sqrt{K}$ to an arbitrary degree zero line bundle $L_z$.
The resulting determinant line bundle is still isomorphic to the Chern-Simons line bundle $(\mathcal{L},h,\nabla)^{d_V}$.

Since the index of $\bar{\partial}_{A,z}^V$ is always zero,
the determinant line bundle admits a canonical section, which can be identified with $\dt_V^{z}\in \mathcal{H}_{d_V}(G)$, up to a phase factor determined by the choice of isometry.

\begin{proposition}\label{regularized det increases}

Along the Yang-Mills flow $A(t)$, regularized determinant $\dt_{\zeta}(\Delta^V_{A(t),z})$ is non decreasing.

\end{proposition}

\begin{proof}

Recall that the squared norm of the canonical section with respect to the Quillen metric 
is given by $\dt_{\zeta}(\Delta^V_{A,z})$. Hence,
\begin{align*}
|\dt_V^z(A)|^2=\dt_{\zeta}(\Delta^V_{A,z}).
\end{align*}
By Lemma \ref{IncreasingNorm}, the squared norm  $|\dt_V^z(A)|^2$ is non decreasing along the Yang-Mills flow $A(t)$.

\end{proof}

\begin{remark}

Essentially, we use the canonical section, which is well-defined only when the index of the family of Dirac operators vanishes.
Hence, the argument extends to a holomorphic vector bundle $V$ over a Riemann surface $\Sigma$ provided its slope satisfies
$\mu(V)=\mathrm{deg}(V)/\mathrm{r}(V)=g-1$.
It can also be generalized to general holomorphic bundles $V$ over higher dimensional K\"ahler manifolds with vanishing Riemann-Roch numbers.

\end{remark}

\subsection{Descriptions of the Unstable Locus}

By Corollary \ref{Vanishing on us},
the canonical sections of determinant line bundles vanish on the unstable locus. We thus obtain a characterization of the unstable locus.

\begin{proposition}

Let $\xi$ be an unstable holomorphic $G_{\mathbb{C}}$-bundle over an elliptic curve $E$, 
and let $V$ be any representation of $G$. Then for every degree-zero line bundle $L_z$ on $E$, we have 
\begin{align*}
H^0(E,\xi(V)\otimes L_z) \neq 0.
\end{align*}

\end{proposition}

\begin{proof}

Let $V$ be a representation of $G$, and let $L_z$ be a degree-zero line bundle over the elliptic curve $E$.
The determinant line bundle is isomorphic to the $d_V$-level of Chern-Simons line bundle over $\mca$.
Moreover, the canonical section $\dt_{V,z}$ of the determinant line bundle vanishes at $A$ if and only if
$H^0(E,V_A\otimes L_z)\neq 0$.

Since $A$ is unstable, Corollary \ref{Vanishing on us} implies that $\dt_{V,z}(A)=0$.
Hence, the corresponding unstable $G_{\C}$-bundle $\xi=P_A$ satisfies
$
H^0(E,\xi(V)\otimes L_z) \neq 0.
$

\end{proof}

This proposition can also be proved directly from the definition of unstable vector bundles on elliptic curves.
Since $\xi(V)$ is an unstable vector bundle of degree zero, there exists a short exact sequence
\begin{align*}
0 \to F \to \xi(V) \to F'\to 0,
\end{align*}
where $F$ is semi-stable and is degree $l>0$. Then we have
\begin{align*}
H^1(E,F\otimes L_0)=H^0(E,(F\otimes L_0)^{\vee})=0.\end{align*}
This follows from the fact that
semi-stable and negative degree vector bundles admit no holomorphic sections \cite[Lemma 17]{Tu1993SemistableBO}.
By Riemann-Roch theorem, $\dim H^0(E,F\otimes L_0)=l$. Since the induced map  $H^0(E,F\otimes L_0) \to H^0(E,\xi(V)\otimes L_0)$ is injective, 
we conclude 
\begin{align*}
\dim H^0(E,\xi(V)\otimes L_0)\geqslant l.
\end{align*}

\subsection{q-Conjugacy Classes of Loop Group}\label{subsec:q-conj class}

In the genus-one case, the moduli stack of holomorphic (algebraic) $G_{\C}$-bundles on the elliptic curve $E_q=\C^*/q^{\Z}$, for $q\in \C^*_{<1}$, is identified with the space of $q$-conjugacy classes in the holomorphic (algebraic) loop group (see \cite{etingof_central_1994,Baranovsky1996ConjugacyCI}).
In this framework, an analog of Theorem \ref{QR=0} can be formulated using trace-class arguments, as developed in \cite{etingof_central_1994,Brchert1998TraceCE}, based on positive-energy representations of the loop group.

Let $LG^h=\mathrm{Hol}(\C^*,G_{\C})$ denote the holomorphic loop group,
which is a complex infinite-dimensional Fr\'echet Lie group.
$\C^*$ acts on $LG^h$ by complex rotation
\begin{align*}
q\cdot g(z):=g(q^{-1}z), q\in \C^*.
\end{align*}
Denote the semi-product $LG\rtimes\C^*$ by
$\widehat{LG^h}$.
Denote the $q$-level set by $LG^h_q=\{(a,q)\in \widehat{LG^h},a\in LG^h\}$. This subset is invariant under the conjugation action of $LG^h$. Indeed, for $g(z)\in LG^h$ and $(a(z),q)\in LG^h_q$, we have $g(z)(a(z),q)g(z)^{-1}=(\prescript{g}{}{a}(z),q)$, where
$
\prescript{g}{}{a}(z)=g(qz)a(z)g(z)^{-1}.
$

Denote by $g: a(z) \mapsto \prescript{g}{}{a}(z)$ the $q$-conjugation action of $LG^h$ on itself, and write $LG^h/_q LG^h$ for the set of $q$-conjugacy classes in $LG^h$.
There is a bijection between $LG^h/_q LG^h$ and the set of holomorphic $G_{\C}$-bundles over the elliptic curve $E_q$.
Given $a\in LG^h$, we define the $G_{\C}$-bundle $P_a\to E_q$ as follows:
\begin{align*}
    P_a=\C^*\times G_{\C}/\sim,
\end{align*}where the equivalence relation $\sim$ is generated by $(z, h) \sim (q z, a(z) h)$.
If $a' = {}^g a$ is a $q$-conjugate of $a$, then the map$(z, h) \longmapsto (z, g(z) h)$
induces an isomorphism of $G_{\mathbb{C}}$-bundles $P_a \xrightarrow{\;\cong\;} P_{a'}$.

We need to clarify the holomorphic structures of $LG^h$. A map $f:S\to LG^h$ is holomorphic if the induced map $\tilde{f}:\C^*\times S\to G_{\C}$ is holomorphic. Similarly, a function
$F:LG^h\to \C$ is holomorphic if $F\circ f:S\to \C$ is holomorphic for any holomorphic map $f:S\to LG^h$ from a finite-dimensional complex manifold $S$.

The universal bundle $\mathcal{P}$ is constructed from the trivial $G_{\C}$-bundle
$\C^*\times G_{\C}\times LG^h$ by imposing the equivalence relation generated by 
\begin{align*}
(z, h, a) \sim (q z, a(z) h, a),
    \quad \text{for all } (z, h, a) \in \mathbb{C}^* \times G_{\mathbb{C}} \times LG^h.
\end{align*}
The $LG^h$-action is given, for $g\in LG^h$, by
\begin{align*}
g\cdot(z,h,a)=(z,g(z)h,\prescript{g}{}{a}).
\end{align*}
This action is compatible with the equivalence relation. Indeed, we have
\begin{align*}
(z,g(z)h,\prescript{g}{}{a})\sim g\cdot(qz,a(z)h,a)=(qz,g(qz)a(z)h,\prescript{g}{}{a})
\sim (qz,\prescript{g}{}{a}(z)g(z)h,\prescript{g}{}{a}).
\end{align*}
Hence, $\mathcal{P}\to E_q\times LG^h$ is a $q$-conjugation equivariant holomorphic principal $G_{\C}$-bundle.
This construction induces a stack morphism $[LG^h/_q LG^h]\to \mathrm{Bund}_{G_{\C}}(E_q)$ by evaluation on $S$-points.

We now construct the Picard generator on $[LG^h/_q LG^h]$ via the central extension of the loop group.
The basic $\C^*$-central extension $\widetilde{LG^h}\to LG^h$ admits a $\C^*$-action induced by complex rotation that lifts the $\C^*$-action on $LG^h$.
Let $\widetilde{LG^h}'=\widetilde{LG^h}\rtimes \C^*\to \widehat{LG^h}$ be the corresponding 
$\C^*$-central extension of the semi-product $\widehat{LG^h}$. This extension is $\widehat{LG^h}$-equivariant. 

Hence, we can restrict the circle bundle to the $q$-level set, and denote the resulting bundle by 
$\widetilde{LG^h_q}\to LG^h_q$. Let $\mathcal{L}^k$ be the associated line bundle over $LG^h_q$ with weight $k$, which is $q$-conjugation equivariant.

\begin{proposition}[{\cite[Lemma 2.3]{Etingof1994SphericalFO}}]

Let $\mathcal{H}$ be a positive-energy representation of $\widetilde{LG^h}$. If $q\in \C_{<1}^*$, then for any $(\widetilde{g},q)\in \widetilde{LG^h_q}$ the operator on a suitable $L^2$-completion of $\mathcal{H}$ induced by 
 $(\widetilde{g},q)$ is trace-class.
We denote its trace by $\Tr_{\mathcal{H}}[(\widetilde{g},q)]$.

\end{proposition}

Given a positive-energy representation $\mathcal{H}$ of level $k$ and $g\in LG^h$, let $\widetilde{g} \in \widetilde{LG^h}$ be a lift of $g$ to the central extension.
If $q\in \C^*_{<1}$, define
\begin{align}
 s_{\mathcal{H}}(g,q):=\big[\widetilde{g},\,\Tr_{\mathcal{H}}((\widetilde{g},q))\big]_k.
\end{align}
The independence of the lift follows from the $\C^*$-equivariance of the trace map $\Tr_{\mathcal{H}}:\widetilde{LG^h_q}\to \C$, which is immediate from the definition of the central extension.
$q$-conjugation equivariance follows from \cite[Proposition 2.5]{Etingof1994SphericalFO}.
Thus 
$s_{\mathcal{H}}\in H_{LG^h}^0(LG^h_q,\mathcal{L}^k)$.

Let $\mathrm{Rep}^k(LG)$ denote the $\C$-vector space spanned by equivalence classes of level-$k$ positive-energy representations.
The assignment $\mathcal{H}\mapsto s_{\mathcal{H}}$ gives a $\C$-linear map
\begin{align}
\rho:\mathrm{Rep}^k(LG)\to H_{LG^h}^0(LG^h_q,\mathcal{L}^k).
\end{align}
Let $T_{\C}\subset LG^h$ denote the subgroup of constant loops taking values in the complex maximal torus. The restricted $q$-conjugation action on $T_{\C}$ coincides with the affine Weyl group action $W_{\mathrm{aff}}=(\Lambda\oplus \tau \Lambda)\rtimes W$ on the Lie algebra $\mathfrak{h}_{\C}$, so that
\[
T_{\C}/(\text{restricted $q$-conj})\;\cong\;\mathfrak{h}_{\C}/W_{\mathrm{aff}}.
\]
Moreover, as in Proposition \ref{Conj into constant connections-}, the $q$-conjugacy classes in $T_{\C}$ parametrize poly-stable $G_{\C}$-bundles over the elliptic curve $E_q$.

Let $\mathrm{ch}:\mathrm{Rep}^k(LG)\to V_k(G)|_{\tau}$ denote the character map sending a representation to its Kac–Weyl character at $\tau \in \mathbb{H}$.
By \cite{Looijenga1976RootSA}, $V_k(G)|_{\tau}$ is spanned by the Kac–Weyl characters of level-$k$ positive-energy representations.
Hence, $\mathrm{ch}$ is an isomorphism.
The following result is given in \cite[Theorem 2.6]{Etingof1994SphericalFO}. To prove injectivity, we again make use of the regular and polystable locus since the stable locus is empty.

\begin{proposition}\label{qr of q-conj}

The following diagram commutes and all maps are isomorphisms of $\C$-vector spaces:
\begin{equation}
 \begin{tikzcd}
\mathrm{Rep}^k(LG) \arrow[rr,"\rho"]\arrow[rd,"\mathrm{ch}"] & & H_{LG^h}^0(LG^h_q,\mathcal{L}^k) \arrow[ld,"\mathrm{r}"] \\
    & V_k(G)|_{\tau}&  
\end{tikzcd}   
\end{equation}

\end{proposition}

\begin{proof}

The commutativity follows from the fact that, when restricted to $T_{\C}$, the trace $\Tr_{\mathcal{H}}$ coincides with the Kac–Weyl character.
We only need to prove the injectivity of $\rho$.
We likewise define the regular and poly-stable locus as the set of elements corresponding to regular and poly-stable $G_{\C}$-bundles over $E_q$.
Arguing as in Proposition \ref{RP Zariski open}, this locus contains a nonempty Zariski open subset of $LG^h$.
It follows that the restriction map
$
H^0_{LG^h}(LG^h_q,\mathcal{L}^k)\to V_k(G)|_{\tau}
$
is injective.
\end{proof}

Proposition \ref{qr of q-conj} can be viewed as an analog of Theorem \ref{QR=0}. However, since the symplectic form on $LG^h$ is not preserved under $q$-conjugation, this is ``quantization commutes with reduction" in the usual symplectic sense.

\bibliographystyle{amsalpha}
\bibliography{references}

@article{Freed1995ClassicalCT,
  title={Classical Chern-Simons Theory, 1},
  author={D. S. Freed},
  journal={Advances in Mathematics},
  year={1995},
  volume={113},
  pages={237-303},
  url={https://api.semanticscholar.org/CorpusID:118559710}
}

@article{SjamarrSlice,
 ISSN = {0003486X, 19398980},
 URL = {http://www.jstor.org/stable/2118628},
 author = {R. Sjamaar},
 journal = {Annals of Mathematics},
 number = {1},
 pages = {87--129},
 publisher = {[Annals of Mathematics, Trustees of Princeton University on Behalf of the Annals of Mathematics, Mathematics Department, Princeton University]},
 title = {Holomorphic Slices, Symplectic Reduction and Multiplicities of Representations},
 urldate = {2025-04-02},
 volume = {141},
 year = {1995}
}

@article{freed1987determinant,
  title={On determinant line bundles},
  author={Freed, D. S.},
  journal={Mathematical aspects of string theory},
  volume={1},
  pages={189--238},
  year={1987},
  publisher={World Scientific}
}

@article{Atiyah1984DiracOC,
  title={Dirac operators coupled to vector potentials.},
  author={M. Atiyah and I. M. Singer},
  journal={Proceedings of the National Academy of Sciences of the United States of America},
  year={1984},
  volume={81 8},
  pages={
          2597-600
        },
  url={https://api.semanticscholar.org/CorpusID:11756103}
}

@article{Guillemin1982GeometricQA,
  title={Geometric quantization and multiplicities of group representations},
  author={V. W. Guillemin and S. Sternberg},
  journal={Inventiones mathematicae},
  year={1982},
  volume={67},
  pages={515-538},
  url={https://api.semanticscholar.org/CorpusID:121632102}
}

@article{Ramanathan1975StablePB,
  title={Stable principal bundles on a compact Riemann surface},
  author={A. Ramanathan},
  journal={Mathematische Annalen},
  year={1975},
  volume={213},
  pages={129-152},
  url={https://api.semanticscholar.org/CorpusID:115307442}
}

@article{Atiyah1983TheYE,
  title={The Yang-Mills equations over Riemann surfaces},
  author={M. Atiyah and R. Bott},
  journal={Philosophical Transactions of the Royal Society of London. Series A, Mathematical and Physical Sciences},
  year={1983},
  volume={308},
  pages={523 - 615},
  url={https://api.semanticscholar.org/CorpusID:13601126}
}

@article{Trautwein2017ASO,
  title={A survey of the GIT picture for the Yang-Mills equation over Riemann surfaces},
  author={S. Trautwein},
  journal={L'Enseignement Mathématique},
  year={2017},
  volume={63},
  pages={63 - 153},
  url={https://api.semanticscholar.org/CorpusID:73622226}
}

@article{Rade1992OnTY,
  title={On the Yang-Mills heat equation in two and three dimensions.},
  author={J. {\r{R}}ade},
  journal={Journal f{\"u}r die reine und angewandte Mathematik (Crelles Journal)},
  year={1992},
  volume={1992},
  pages={123 - 164},
  url={https://api.semanticscholar.org/CorpusID:118058577}
}

@article{Teleman1998TheQC,
  title={The quantization conjecture revisited},
  author={C. Teleman},
  journal={Annals of Mathematics},
  year={1998},
  volume={152},
  pages={1-43},
  url={https://api.semanticscholar.org/CorpusID:17691297}
}

@article{Axelrod1991GeometricQO,
  title={Geometric quantization of Chern-Simons gauge theory},
  author={S. Axelrod and S. D. Pietra and E. Witten},
  journal={Journal of Differential Geometry},
  year={1991},
  volume={33},
  pages={787-902},
  url={https://api.semanticscholar.org/CorpusID:119974874}
}

@article{Friedman1997PrincipalGB,
  title={Principal G bundles over elliptic curves},
  author={R. Friedman and J. W. Morgan and E. Witten},
  journal={Mathematical Research Letters},
  year={1997},
  volume={5},
  pages={97-118},
  url={https://api.semanticscholar.org/CorpusID:14347135}
}

@article{Freed:2016mpb,
    author = "Freed, D. S.",
    title = "{On equivariant Chern{\textendash}Weil forms and determinant lines}",
    eprint = "1606.01129",
    archivePrefix = "arXiv",
    primaryClass = "math.DG",
    doi = "10.4310/sdg.2017.v22.n1.a5",
    journal = "Surveys Diff. Geom.",
    volume = "22",
    number = "1",
    pages = "125--132",
    year = "2017"
}

@article{Quillen1985DeterminantsOC,
  title={Determinants of Cauchy-Riemann operators over a Riemann surface},
  author={D. Quillen},
  journal={Functional Analysis and Its Applications},
  year={1985},
  volume={19},
  pages={31-34},
  url={https://api.semanticscholar.org/CorpusID:122340883}
}

@article{Bismut1986TheAO,
  title={The analysis of elliptic families. I. Metrics and connections on determinant bundles},
  author={J.-M. Bismut and D. S. Freed},
  journal={Communications in Mathematical Physics},
  year={1986},
  volume={106},
  pages={159-176},
  url={https://api.semanticscholar.org/CorpusID:55389271}
}

@article{Ramadas1989SomeCO,
  title={Some comments on Chern-Simons gauge theory},
  author={T. R. Ramadas and I. M. Singer and Jonathan Weitsman},
  journal={Communications in Mathematical Physics},
  year={1989},
  volume={126},
  pages={409-420},
  url={https://api.semanticscholar.org/CorpusID:119716826}
}

@article{Laszlo1997TheLB,
  title={The line bundles on the moduli of parabolic G-bundles over curves and their sections},
  author={Y. Laszlo and C. Sorger},
  journal={Annales Scientifiques De L Ecole Normale Superieure},
  year={1997},
  volume={30},
  pages={499-525},
  url={https://api.semanticscholar.org/CorpusID:9613079}
}

@article{Freed1986DeterminantsTA,
  title={Determinants, torsion, and strings},
  author={D. S. Freed},
  journal={Communications in Mathematical Physics},
  year={1986},
  volume={107},
  pages={483-513},
  url={https://api.semanticscholar.org/CorpusID:121998986}
}

@article{Bismut2,
author = {J.-M. Bismut and D. S. Freed},
title = {{The analysis of elliptic families. II. Dirac operators, eta invariants, and the holonomy theorem}},
volume = {107},
journal = {Communications in Mathematical Physics},
number = {1},
publisher = {Springer},
pages = {103 -- 163},
year = {1986},
}

@article{Looijenga1976RootSA,
  title={Root systems and elliptic curves},
  author={E. Looijenga},
  journal={Inventiones mathematicae},
  year={1976},
  volume={38},
  pages={17-32},
  url={https://api.semanticscholar.org/CorpusID:121219851}
}

@article{Bismut1999SymplecticGA,
  title={Symplectic geometry and the Verlinde Formulas},
  author={J.-M. Bismut and F. Labourie},
  journal={Surveys in differential geometry},
  year={1999},
  volume={5},
  pages={97-311},
  url={https://api.semanticscholar.org/CorpusID:121272467}
}

@article{Meinrenken1997SINGULARRA,
  title={SINGULAR REDUCTION AND QUANTIZATION},
  author={E. Meinrenken and R. Sjamaar},
  journal={Topology},
  year={1997},
  volume={38},
  pages={699-762},
  url={https://api.semanticscholar.org/CorpusID:18573269}
}

@article{Atiyah1971RiemannSA,
  title={Riemann surfaces and spin structures},
  author={M. Atiyah},
  journal={Annales Scientifiques De L Ecole Normale Superieure},
  year={1971},
  volume={4},
  pages={47-62},
  url={https://api.semanticscholar.org/CorpusID:55373563}
}

@article{Gawdzki1994SU2WT,
  title={SU(2) WZW theory at higher genera},
  author={K. Gaw{\c{e}}dzki},
  journal={Communications in Mathematical Physics},
  year={1994},
  volume={169},
  pages={329-371},
  url={https://api.semanticscholar.org/CorpusID:119514829}
}

@article{Beauville1993ConformalBA,
  title={Conformal blocks and generalized theta functions},
  author={A. Beauville and Y. Laszlo},
  journal={Communications in Mathematical Physics},
  year={1993},
  volume={164},
  pages={385-419},
  url={https://api.semanticscholar.org/CorpusID:15034616}
}

@article{Tian1998AnAP,
  title={An analytic proof of the geometric quantization conjecture of Guillemin-Sternberg},
  author={Y. Tian and W. Zhang},
  journal={Inventiones mathematicae},
  year={1998},
  volume={132},
  pages={229-259},
  url={https://api.semanticscholar.org/CorpusID:119943992}
}

@article{Brchert1998TraceCE,
  title={Trace Class Elements and Cross-Sections in Kac-Moody Groups},
  author={G. Br{\"u}chert},
  journal={Canadian Journal of Mathematics},
  year={1998},
  volume={50},
  pages={972 - 1006},
  url={https://api.semanticscholar.org/CorpusID:124433932}
}

@misc{Friedman1998HolomorphicPB,
  title={Holomorphic principal bundles over elliptic curves},
  author={R. Friedman and J. W. Morgan},
publisher={arxiv},
  year={1998},
  url={arXiv:math/9811130}
}

@article{Etingof1994SphericalFO,
  title={Spherical functions on affine Lie groups},
  author={P. Etingof and I. Frenkel and A. Kirillov},
  journal={Duke Mathematical Journal},
  year={1994},
  volume={80},
  pages={59-90},
  url={https://api.semanticscholar.org/CorpusID:16800000}
}

@misc{witten_verlinde_1993,
    title = {The Verlinde Algebra And The Cohomology Of The Grassmannian},
    url = {http://arxiv.org/abs/hep-th/9312104},
    language = {en},
    urldate = {2024-01-31},
    publisher = {arXiv},
    author = {Witten, E.},
    month = dec,
    year = {1993},
    note = {arXiv:hep-th/9312104},
}

@article {Baranovsky1996ConjugacyCI,
    AUTHOR = {Baranovsky, V. and Ginzburg, V.},
     TITLE = {Conjugacy classes in loop groups and {$G$}-bundles on elliptic
              curves},
   JOURNAL = {Internat. Math. Res. Notices},
  FJOURNAL = {International Mathematics Research Notices},
      YEAR = {1996},
    NUMBER = {15},
     PAGES = {733--751},
      ISSN = {1073-7928,1687-0247},
   MRCLASS = {20G35 (14F05)},
  MRNUMBER = {1413870},
MRREVIEWER = {Vladimir\ L.\ Popov},
       DOI = {10.1155/S1073792896000463},
       URL = {https://doi.org/10.1155/S1073792896000463},
}

@article{etingof_central_1994,
  title={Central extensions of current groups in two dimensions},
  author={Etingof, P. and Frenkel, I.},
  journal={Communications in Mathematical Physics},
  volume={165},
  number={3},
  pages={429--444},
  year={1994},
  publisher={Springer}
}

@article{Meinrenken1995SymplecticSA,
  title={Symplectic Surgery and the Spinc–Dirac Operator},
  author={E. Meinrenken},
  journal={Advances in Mathematics},
  year={1995},
  volume={134},
  pages={240-277},
  url={https://api.semanticscholar.org/CorpusID:50262632}
}

@article{Kumar1994InfiniteGA,
  title={Infinite Grassmannians and moduli spaces ofG-bundles},
  author={S. Kumar and M. S. Narasimhan and A. Ramanathan},
  journal={Mathematische Annalen},
  year={1994},
  volume={300},
  pages={41-75},
  url={https://api.semanticscholar.org/CorpusID:119570247}
}

@article{Ma2008GeometricQF,
  title={Geometric quantization for proper moment maps: the Vergne conjecture},
  author={X. Ma and W. Zhang},
  journal={Acta Mathematica},
  year={2008},
  volume={212},
  pages={11-57},
  url={https://api.semanticscholar.org/CorpusID:212105}
}

@article{Huebschmann1995TheSO,
  title={The singularities of Yang-Mills connections for bundles on a surface},
  author={J. Huebschmann},
  journal={Mathematische Zeitschrift},
  year={1995},
  volume={220},
  pages={595-609},
  url={https://api.semanticscholar.org/CorpusID:15274907}
}

@article{Tu1993SemistableBO,
  title={Semistable Bundles over an Elliptic Curve},
  author={L. W. Tu},
  journal={Advances in Mathematics},
  year={1993},
  volume={98},
  pages={1-26},
  url={https://api.semanticscholar.org/CorpusID:119468636}
}
\end{document}